\documentclass[preprint,3p]{elsarticle}

\usepackage{adjustbox}
\usepackage{graphicx}
\usepackage{wrapfig}
\usepackage{epsfig}
\usepackage{booktabs}
\usepackage[usenames,dvipsnames]{color}
\usepackage{multirow}
\usepackage{amsfonts,amsmath,amssymb}
\usepackage{subfig}
\usepackage{textcomp}
\usepackage{mathrsfs}
\usepackage{amsthm}

\usepackage{amsmath}

\usepackage{bbm}

\usepackage{yfonts}

\usepackage[utf8]{inputenc}

\usepackage[boxed]{algorithm2e}
\usepackage{mathrsfs}
\usepackage{lineno} \modulolinenumbers[1]
\usepackage{hyperref}
\usepackage{enumitem}

\usepackage{caption}
\def\beq{\begin{equation}}
\def\eeq{\end{equation}}

\def\XXint#1#2#3{{\setbox0=\hbox{$#1{#2#3}{\int}$ }
\vcenter{\hbox{$#2#3$ }}\kern-.55\wd0}}

\def\epsilon{\varepsilon}

\def\beq{\begin{equation}}
\def\eeq{\end{equation}}
\def\beqa{\begin{eqnarray}}
\def\eeqa{\end{eqnarray}}

\journal{COST}

\begin{document}

\begin{frontmatter}

\title{Numerical study of interface cracking in composite structures using a novel geometrically nonlinear Linear Elastic Brittle Interface Model: mixed-mode fracture conditions and application to structured interfaces}

\author[1]{L. Garc\'{i}a-Guzm\'{a}n}
\author[1]{J. Reinoso  \corref{cor1}}
\ead{jreinoso@us.es}
\cortext[cor1]{Corresponding author. Tel.:+34 954487299; Fax: +34 954461637}
\author[1,2]{A. Valderde}
\author[3]{E. Mart\'{i}nez-Pañeda}
\author[1]{L. T\'{a}vara}

\address[1]{Grupo de Elasticidad y Resistencia de Materiales, \\ Escuela T\'{e}cnica Superior de Ingenier\'{\i}a, Universidad de Sevilla, \\ Camino de los Descubrimientos s/n, 41092 Sevilla, Spain}
\address[2]{IMT School for Advanced Studies, Piazza San Francesco 19, Lucca 55100, Italy}
\address[3]{Department of Civil and Environmental Engineering, Imperial College London, London SW7 2AZ, UK}

\begin{abstract}
Interface cracking is one of the most prominent failure modes in fibre reinforced polymer (FRP) composites. Recent trends in high-tech applications of FRP composites exploit the limits of the load bearing capacity, generally encompassing the  development of notable nonlinear effects from geometrical and material signatures. In this investigation, we present a comprehensive assessment of the  new Linear Elastic Brittle Interface Model (LEBIM) in geometrically nonlinear applications undergoing mixed-mode fracture conditions. This interface model for triggering fracture events is formulated through the advocation of   continuum-like assumptions (for initial non-zero interface thickness) and allows the incorporation of the potential role of the in-plane deformation effects. The performance of the present interface model is demonstrated through the simulation of specimens with mixed-mode delamination, with special attention for its application in samples equipped with structured interfaces. Current predictions exhibit  an excellent agreement with respect to experimental data, validating the proposed methodology.   
\end{abstract}


\end{frontmatter}

\section{Introduction}

The recurrent  requirements for the achievement of high strength-to-weight ratios in different engineering applications have  led to the  continuous improvement    of production techniques and methodologies  of analysis.  In this direction,    fibre reinforced polymers (FRP) composite materials have become particularly popular relative to conventional materials (especially in contrast to metals) due to their appealing strength and stiffness properties, widening the current ranges of applicability within  the   aerospace, automotive or renewable industries, among other sectors.

However, the inherent heterogeneous character of FRP composites at several scales of observation  entails characteristic failure phenomena between the composing entities and constituents. This is the case, for instance, of delamination events at the macro-scale  \cite{Turon06,camanhodavila2003,TAVARA2019362} and fibre-matrix debonding \cite{TAVARA2011207,Zumaquero18} at the micro-scale, among many other debonding-like failures in FRP composites. Such cracking events can be principally caused either  by external loading actions  or induced by manufacturing and joining processes   \cite{Banea09}. Motivated by these failure  phenomena,  significant research efforts  have been conducted in recent years towards  the efficient  incorporation of  alternative joining procedures; such as adhesive bonding, a compelling technique that provides additional advantages in terms of the mechanical responses in conjunction with  the enhancement of fatigue and environmental performances \cite{reinoso2016damage}.

The understanding of failure mechanisms in solids,  with special interest on joints/interfaces,  has been of  high interest in both industrial and research contexts,  striving for  the development of different prediction methodologies. Thus, on the one hand, the Linear Elastic  Fracture Mechanics (LEFM) approach, relying on  its energetic version, makes use of an energy criterion  to predict the failure either in adherents, adhesive or the interface between them. The energy-based LEFM was originally proposed by Griffith \cite{Griffith1921} and posteriorly  revisited by Irwin \cite{Irwin57}. One of the most popular LEFM-based methodologies is the so-called    Virtual Crack Closure Technique (VCCT), where the crack advance is triggered as long as  the energy release rate exceeds a certain threshold or critical value under pure or mixed-mode fracture conditions \cite{HUTCHINSON199163}. In this regard, studies of the stress intensity factors for homogeneous and multi-material specimens have been comprehensively addressed in \cite{Barroso2003,BARROSO20082671,XU1999775,GROTH1988107} in order to determine proper conditions for fracture progression, whereas the  extension on the  application of Fracture Mechanics to nonlinear  materials was conducted by Rice and co-authors  through the so-called J-Integral method \cite{Rice68}.

Alternatively to fracture mechanics-based methods,  a different perspective for predicting  fracture response in solids  can be  advocated  by means of Cohesive Zone Model (CZM) techniques, which have been extensively used for triggering interface fracture phenomena \cite{Reinoso2014}. Originally proposed by Barenblatt \cite{BARENBLATT196255} and Dugdale \cite{dugdale1960}, CZMs are generally formulated within the context of Damage Mechanics of irreversible processes \cite{REINOSO2017116,holz}. Thus, in the particular case of interface  fracture,  the scalar-based damage variable  within the CZ formulation accounts for the stiffness degradation within the so-called fracture process zone (FPZ) obeying the particular form of the  traction-separation law (TSL). This  TSL  relates the displacement jumps across the interface with the respective traction components \cite{Wang13}. The flexibility of CZMs  in terms of the TSL definition (featuring bilinear \cite{GEUBELLE1998589}, trapezoidal \cite{TVERGAARD93}, exponential \cite{Ortiz99} laws, among many others) permits the characterization of different adhesives or interfaces, whilst the proper parameters of  the cohesive zone can be extracted  from experimental data as proposed in  \cite{SORENSEN20031841,maloney18}.

An interesting approach in interface fracture mechanics is endowed through the consideration of the interface/joint as a continuous distribution of linear springs. This interface formulation, usually denominated as Linear Elastic-Brittle Interface Model (LEBIM), encompasses a  linear elastic relationship between the displacement jumps and the corresponding  tractions  across the interface up to the abrupt failure, that is tracked  once a particular fracture criterion is violated.   This methodology was proposed by Prandtl \cite{Prandtl2011} and Mott \cite{Mott48}, and it is mainly  suitable for scenarios where the overall stiffness is  ruled by the adherents, and the  gradual  stiffness loss due to the interface degradation can be neglected.   Thus, this interface conception has been efficiently  used to represent the behaviour of  brittle-like  interfaces, such as epoxy-based adhesives \cite{Mantic,TAVARA2017148,TAVARA2019362,reinoso2016damage}.

Another key aspect  for accurate predictions   within the context of interface/joint fracture concerns the rigorous  selection of  the kinematic  hypotheses in accordance with the experimental conditions, that is, whether  material and geometrical nonlinearities concomitantly evolve throughout the  numerical   analysis.  This aspect has a direct reflection on the way through which the stress and strain fields are computed, and therefore determining the onset and  propagation of   failure.   In many lightweight structures, such as those  extensively used in aerospace or renewable industry (stiffened panels, turbine blades, among many others),   high-performance materials permit the evolution of large displacements during the loading  applications    prior reaching the corresponding collapsing points. Principal after-effect is the non-negligible difference between original and current configurations, that leads to inaccurate calculations if the analysis is restricted to  small-displacement theory. 
These considerations were comprehensively analysed for cohesive-like interfaces in  \cite{REINOSO201737,REINOSO2017116,Reinoso2014,Paggi2015}. In this concern, Ortiz and Pandolfi \cite{Ortiz99} proposed a surface-like finite elements in which the normal and tangential directions to the surface are monitored, and every geometrical operation is carried out on the middle surface of the element. This procedure allows superimposed rigid motions to be overcome. Alternatively, Qiu. et al \cite{QIU20011755} applied a simple corotational formulation to one-dimensional interface elements, whereas Reinoso and Paggi developed  2D \cite{REINOSO2017116} and 3D \cite{PAGGI2015106} interface elements for large deformation analysis dealing with  geometrical and material nonlinearities using a consistent derivation of the corresponding operators.

\begin{figure}[h!]
	\centering
	\includegraphics[width=0.8\textwidth]{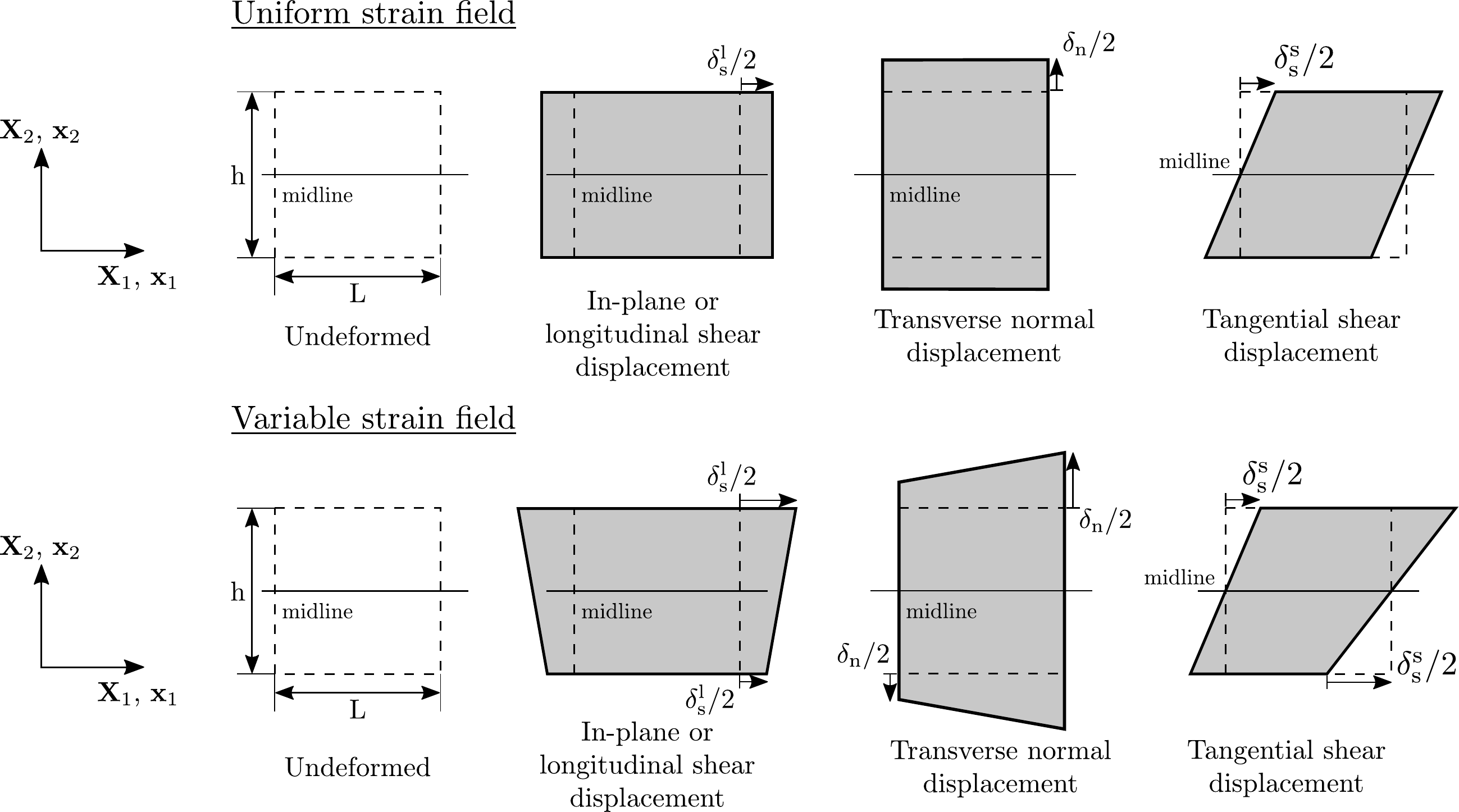}
	\caption{Displacement field within a 2D element in the interface model under uniform and variable strain field: transverse normal $\delta_\text{n}$, longitudinal shear or in-plane $\delta_{\text{s}}^{\text{l}}$ and tangential shear $\delta_\text{s}^\text{s}$.}
	\label{TSL_disp}
\end{figure}

Differing from precedent methodologies for triggering interface cracking that incorporate   geometrically nonlinear effects, the authors proposed an   alternative formulation  \cite{GarciaGuzman20}, that is denominated a ``consistent finite displacement and rotation formulation of the Linear Elastic Brittle Interface Model''. This interface model   can be easily integrated within standard continuum finite elements as user-defined material subroutine and accounts for the potential effects of in-plane or longitudinal normal deformations (variations along the bondline direction).  Complying with such modelling technique, the separation between top and bottom surfaces respect to the interface midplane can be  determined through the deformation gradient $\mathbf{F}$ under large displacement conditions. This standpoint presents some advantages over other methods, such as its simplicity (it is not required the coding of a new element formulation) and the computation of a complete displacement field including transverse normal gap $\delta_\text{n}$, tangential shear gap $\delta_\text{s}^\text{s}$ and longitudinal shear or in-plane gap $\delta_{\text{s}}^{\text{l}}$, as shown in Fig.~\ref{TSL_disp}. Thus, recalling  the predictions presented in \cite{GarciaGuzman20}, the new geometrically  nonlinear LEBIM  formulation does offer very promising results and notably simplifies the implementation  requirements. Within this context, the principal objective of the current investigation is the comprehensive validation of the   geometrically nonlinear LEBIM  \cite{GarciaGuzman20} for mixed-mode loading conditions  and for its usage in structured interfaces as in \cite{GarciaGuzman19,garcia-guzman}.

The organization of the manuscript is as follows. Section \ref{interface} outlines an overview of the interface modeling. Validation of the current interface model through its assessment  for  Double Cantilever Beam (DCB), Mixed Mode Bending (MMB) and End Notch Flexure (ENF) tests is detailed in Section \ref{app}, whereas its application to structured interfaces is presented  in Section \ref{Trap}. Finally, the main conclusions of this investigation are summarized  in Section \ref{section4}.

\section{Interface modelling: general aspects and formulation}
\label{interface}

%
\subsection{Geometrically nonlinear interface model}
\label{large_disp}

This Section presents the main aspects of the new LEBIM formulation for geometrically nonlinear applications. The current formulation  is compatible with general-purpose solid elements and it is subsequently particularized for 2D analysis. Complying with a finite thickness interface model that can be integrated into standard continuum finite elements, the required displacements   for the evaluation of the LEBIM   traction-separation law, i.e. the relative transverse normal displacement $\delta_\text{n}$, the tangential shear displacement $\delta_{\text{s}}^{\text{s}}$ and the longitudinal shear or in-plane displacement $\delta_{\text{s}}^{\text{l}}$, are referred to the element midline.  Thus, the tracking of this midline can be  performed using material user-subroutine \texttt{UMAT} supported by the commercial software \texttt{ABAQUS}{\textsuperscript{\textregistered}} \cite{abaqus}. See a comprehensive description of the computations at the material point level in \cite{GarciaGuzman20}.  

Assuming the finite displacement theory \cite{bonet_wood_2008}, the deformation gradient $\mathbf{F}$ is a two-point tensor that relates current $\mathbf{x}$ and initial $\mathbf{X}$ configurations, considering deformations as well as rigid body motions. In this modelling framework, two different standpoints can be adopted: Lagrangian or material description, in which the variables are referred to the initial configuration, or Eulerian or spatial description, in which the variables are referred to the current configuration according to:

\begin{equation}
\label{F}
\mathbf{F} = \dfrac{\partial \mathbf{x}}{\partial \mathbf{X}} = 
\left[ \begin{array}{cc}  
F_{11} & F_{12} \\ 
F_{21} & F_{22}
\end{array} \right].
\end{equation}

\begin{figure}[h!]
	\centering
	\includegraphics[width =0.5 \textwidth]{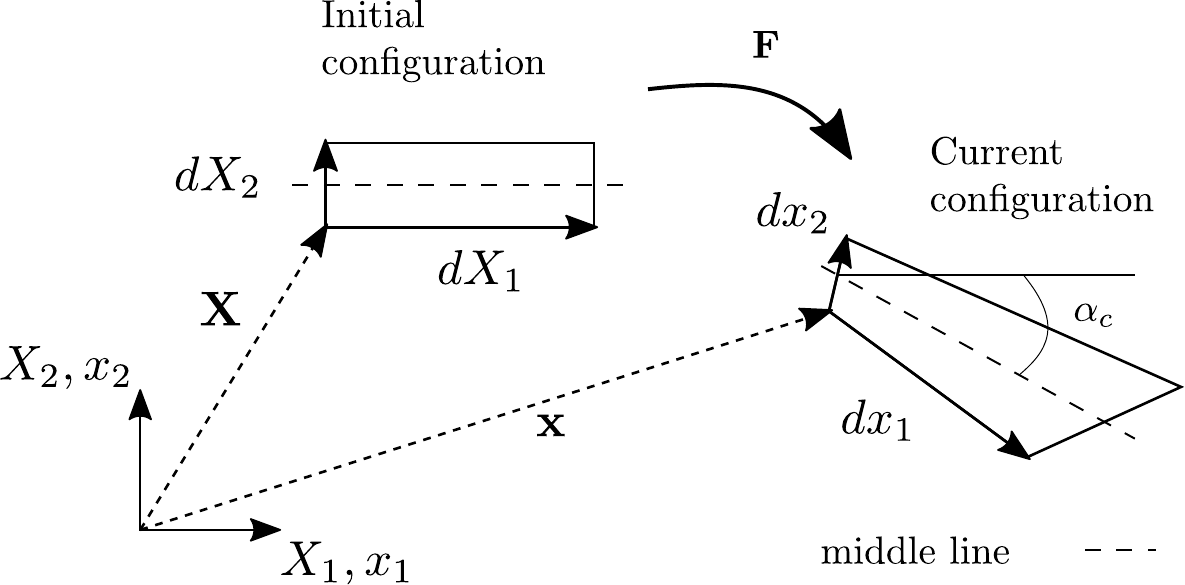}
	\caption{Initial and current configurations of an element and its relation through the deformation gradient tensor $\mathbf{F}$.}
	\label{UMAT}
\end{figure}

The computation of the polar decomposition allows the deformation gradient $\mathbf{F}$ to be split into the stretch and rotation tensors. As recalled in \cite{GarciaGuzman20}, this operation can be executed in two ways: (i) the application of the deformation is applied first, and then  rotation (material description, Eq. \eqref{polar1}), or conversely (ii)  the application  of the   rotation is  followed by the insertion of the  deformation at the material point level (spatial description, Eq. \eqref{polar2}), as
\begin{equation}
\label{polar1}
\mathbf{F} = \mathbf{RU},
\end{equation}

\begin{equation}
\label{polar2}
\mathbf{F} = \mathbf{VR},
\end{equation}
\noindent
where $\mathbf{R}$ is the rotation tensor, $\mathbf{U}$ is the stretch tensor according to a material description and $\mathbf{V}$ is the stretch tensor following a spatial description. A graphical representation of the motion considering finite strain theory is depicted  in Fig.~\ref{UMAT}.

For the sake of simplicity, the former expression is used within this study, adopting a Lagrangian standpoint. The computation of the element midline rotation can be obtained through the following expression

\begin{equation}
\label{alpha}
\tan (\alpha_{\text{c}}) = \frac{F_{21}}{F_{11}},
\end{equation}
\noindent
where the significance of the angle  $\alpha_{\text{c}}$ is shown in Fig.~\ref{UMAT}. Accordingly, the rotation and stretch tensors, $\mathbf{R}$ and $\mathbf{U}$, can be easily calculated as

\begin{equation}
\label{rotation}
\mathbf{R} = \left[ \begin{array}{cc}  
\cos(\alpha_{\text{c}}) & -\sin(\alpha_{\text{c}}) \\ 
\sin(\alpha_{\text{c}}) & \cos(\alpha_{\text{c}})
\end{array} \right],
\end{equation}

\begin{equation}
\label{U_alpha0}
\mathbf{U} = \mathbf{R}^{\text{T}} \mathbf{F} = \left[ \begin{array}{cc}  
U_{11} & U_{12} \\ 
0 & U_{22}
\end{array} \right] .
\end{equation}

Thus, the deformation tensor can be obtained in a straightforward manner through taking into account the rotation of the midline element. Further details about the procedure to obtain $\alpha_c$, components of the $\mathbf{U}$ tensor and differences with respect to regular continuum elements can be found in \cite{GarciaGuzman20}. The next ingredient for the evaluation of the TSL is the computation of the displacement jumps across the interface:   $\delta_\text{n}$, $\delta_\text{s}^\text{s}$ and $\delta_{\text{s}}^{\text{l}}$.
The computation of the  displacement vector can be recalled via the definition of the variation of the displacement field. This expression  shows the relation between the deformation gradient tensor $\mathbf{F}$ and undeformed vectors $\textrm{d}\mathbf{X}$ as

\begin{equation}
\label{relative_disp}
\textrm{d}\mathbf{x} = \textrm{d}\mathbf{X} + \textrm{d}\mathbf{u} \Longrightarrow \textrm{d}\mathbf{u} = \textrm{d}\mathbf{x}-\textrm{d}\mathbf{X} = (\mathbf{F}-\mathbf{I}) \textrm{d}\mathbf{X}.
\end{equation}
\noindent
In absence of a rigid body rotation, the tensor $\mathbf{R}$ is equal to the identity matrix and the relative displacements can be expressed as
\begin{equation}
\label{relative_disp2}
d\mathbf{u} = (\mathbf{U}-\mathbf{I}) \textrm{d}\mathbf{X}.
\end{equation}
\noindent
From a different perspective, in terms of a generalised material strain tensor of order $n$,  the corresponding strain tensors can be computed as

\begin{equation}
\label{gen_strain}
\mathbf{E}^{n} = \frac{1}{n}(\mathbf{U}^n-\mathbf{I}),
\end{equation}

\begin{equation}
\label{gen_strain1}
\mathbf{E}^{1} = (\mathbf{U}-\mathbf{I}).
\end{equation}

The operator $(\mathbf{U}-\mathbf{I})$ can be seen as the material strain tensor of order $n=1$ (Eq. \ref{gen_strain1}) or the so-called Biot strain tensor.  Restricting our attention to the definition of the first order strain tensor, one obtains

\begin{equation}
\label{gen_strain2}
\mathbf{E}^{1}  = \left[ \begin{array}{cc}  
U_{11}-1 & U_{12} \\ 
0 & U_{22}-1
\end{array} \right]  = 
\left[ \begin{array}{cc}  
\frac{\partial \delta_{\text{s}}^{\text{l}}}{\partial X_1} & \frac{\partial \delta_\text{s}^\text{s}}{\partial X_2} \\ 
0 & \frac{\partial \delta_\text{n}}{\partial X_2}
\end{array} \right].
\end{equation}
\noindent
Through the proper selection of the  undeformed element dimensions, L and h, as the initial vectors $\textrm{d} X_1$ and $\textrm{d} X_2$ respectively, the displacement  jumps can be computed as follows

\begin{equation}
\label{delta_1}
\delta_{\text{s}}^{\text{l}}   = (U_{11}-1)\textrm{d}X_1 = (U_{11}-1) \text{L},
\end{equation}

\begin{equation}
\label{delta_n}
\delta_\text{n}   = (U_{22}-1)\textrm{d}X_2 = (U_{22}-1) \text{h} ,
\end{equation}

\begin{equation}
\label{delta_s}
\delta_\text{s}^\text{s}  = U_{12}\textrm{d}X_2  = U_{12} \text{h} .
\end{equation}

In applications experiencing rigid body rotations, the  motion can be described as follows: firstly, the element is deformed (material description) through the material stretch tensor $\mathbf{U}$ and the gap displacements or separations $\delta_\text{n}$, $\delta_\text{s}^\text{s}$ and $\delta_{\text{s}}^{\text{l}}$ are determined; secondly, the element is rotated via the tensor $\mathbf{R}$ in order to get the current position.

Regarding, the extension of this procedure to 3D applications would require the calculation of three angles in order to track the motion of the element's middle plane during the analysis. That is, the position of the current axis (placed on the element's midplane) with respect to the reference configuration is described by three rotations. This fact increases the complexity of the procedure in comparison to 2D analysis, where only a single angle is needed to characterise the midline behaviour. Nevertheless, analogue procedures to the 2D analysis applied to the different directions will lead to the rotation matrix and the displacement field in 3D scenarios. Regarding LEBIM, 3D proposals for small displacement scenarios have been recently used for some problems including composite laminates \cite{TAVARA2019362,reinoso2016damage}.

\subsection{Constitutive interface equations: Linear Elastic Brittle Interface Model}
\label{Contitutibe_LEBIM}

Once the separation  gaps at an interface are computed,  the next phase requires the determination of the traction vector for the evaluation of the TSL that characterizes the interface failure.   As stated above, in the related literature there are a wide  variety of traction-separation laws (bilinear, exponential, trapezoidal, etc) enabling the characterization of  different interface behaviours (ductile, brittle, etc). In this investigation the Linear Elastic Brittle Interface Model (LEBIM) is employed in the subsequent applications, in which the traction  and energy standpoints merge in an unique criterion. Nevertheless, it is worth mentioning that the current methodology can be applicable to any different profile of TSL without remarkable limitations.  

In a general sense, the LEBIM is herewith used to characterize the interface between two solids or a thin adhesive layer, whose stress profile along the thickness is uniform. This technique can be conceived as an elastic spring foundation with a cut-off traction response, that features the abrupt failure \cite{Mantic,TAVARA2019362}. 
Accordingly, energy dissipation before crack propagation is considered as  negligible and, therefore, no softening area ahead of the crack tip should be appreciated in the problem. Additionally, as   was observed in \cite{TAVARA2019362,reinoso2016damage}, if the stiffness of the system is mostly governed by the adherents, the shape of the TSL has a minor influence on  the corresponding overall results, i.e. classical cohesive zone models (with a large softening zone) will lead to similar results as those obtained by LEBIM.

\begin{figure}[h!]
	\centering
	\includegraphics[width =0.65 \textwidth]{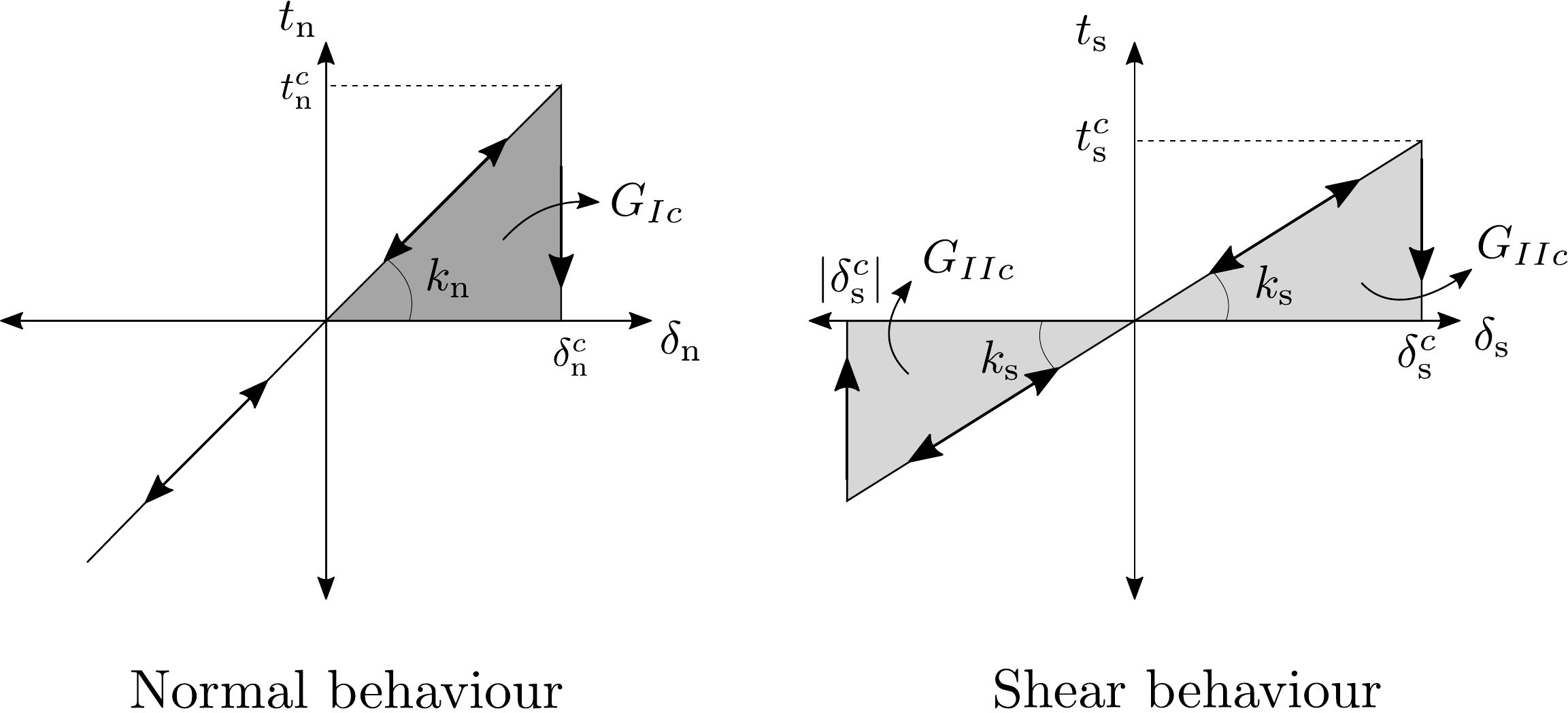}
	\caption{Linear Elastic Brittle Interface Model: traction-separation law in normal ($t_{\text{n}} - \delta_\text{n}$) and shear ($t_{\text{s}} - \delta_\text{s}$) directions.}
	\label{LEBIM}
\end{figure}

Specifically, tractions $t$ and energy stored $G$ in this ``spring distribution'' are used to compute the mixed mode ratio $B  = G_{II}/G_{T}$, where $G_{T} = \left( G_{I} + G_{II}\right)$, the critical fracture energy $G_c$ and the failure instant. Following \cite{Mantic}, let tractions be described in terms of  displacement jumps across the interface  by means of a linear elastic law as

\begin{equation}
\label{tn}
\begin{aligned}
t_n =
\begin{cases}
k_{\text{n}} \delta_\text{n},    & \quad \text{if } \delta_\text{n}  \leq \delta_\text{n}^c \\
0, & \quad \text{otherwise, }
\end{cases}
\end{aligned},
\end{equation}

\begin{equation}
\label{ts}
\begin{aligned}
t_{\text{s}} =
\begin{cases}
k_{\text{s}} \delta_\text{s}^\text{s} + k_{\text{s}} \delta_\text{s}^\text{l} = k_{\text{s}} \delta_\text{s},    & \quad \text{if } \left| \delta_{\text{s}}^{\text{s}} \right|  \leq \left| \delta_{\text{n}}^c \right| \\
0, & \quad \text{otherwise, }
\end{cases}
\end{aligned},
\end{equation}
\noindent
where $t_{\text{n}}$ and $t_{\text{s}}$ are normal and shear tractions, $\delta_\text{n}$, $\delta_{\text{s}}^{\text{s}}$ and $\delta_{\text{s}}^{\text{l}}$ are relative transverse normal, tangential shear and in-plane displacements and $k_{\text{n}}$ and $k_{\text{s}}$ are normal and shear stiffnesses, respectively. Note that the shear jump $\delta_\text{s}$ admits two contributions $\delta_\text{s} = \delta_\text{s}^\text{s} + \delta_\text{s}^\text{l} $ and that $k_{\text{n}}$ and $k_{\text{s}}$ are expressed in [$\frac{\text{MPa}}{\text{mm}}$]. Fig.~\ref{LEBIM} depicts the behaviour of LEBIM  constitutive law.

Notwithstanding, tractions drop to zero when $t^c$ is reached or, in energy terms, if the energy stored $G$ fulfills the fracture toughness criterion: $G = G_c$. Hence, the definition of the energy release rate stored $G$ and fracture toughness $G_c$ is fundamental in the failure description. In this way, the energy contributions can be split  into those associated with fracture Mode I and Mode II counterparts ($G_I$ and $G_{II}$ respectively) for mixity purposes, which respectively correspond to normal and shear behaviour according to the expressions:

\begin{equation}
\label{G}
G = G_{I}+G_{II},
\end{equation}

\begin{equation}
\label{G_I}
G_I = \frac{\left< t_{\text{n}} \right>_+\left< \delta_\text{n} \right>_+}{2} = \frac{\left< t_{\text{n}} \right>_+^2}{2 k_{\text{n}}},
\end{equation}

\begin{equation}
\label{G_II}
G_{II} = \frac{ t_{\text{s}}  \delta_\text{s} }{2} = \frac{ t_{\text{s}} ^2}{2 k_{\text{s}}},
\end{equation}
\noindent
where $G_{I}$ and $G_{II}$ are the energy release rates for fracture Mode I and Mode II, respectively. In the previous expressions, the symbol $\left< \right>$ stands for the Macaulay brackets, and therefore only positive values of normal tractions and displacements are used for  the $G_{I}$ calculation. 

Finally, a critical fracture energy criterion involving any mixed mode condition $G_c(B)$ establishes the limit condition. Without any loss of generality, we advocate in the present investigation the use of the phenomenological  Benzeggah-Kenane (BK) criterion \cite{BK1996}, whose mathematical expression renders

\begin{equation}
\label{Gc}
G_{c} = G_{Ic}+\left( G_{IIc} - G_{Ic} \right)\left( \frac{G_{II}}{G_{I}+G_{II}} \right)^{\eta},
\end{equation}
\noindent
where $\eta$ is a material coefficient as described in \cite{Camanho_delamination}.

\subsection{Snap-back control algorithm}
\label{snapback}

From the numerical point of view, in simulations involving damage progression,  the nonlinear effects     play an important role in the analysis convergence in terms of achieving equilibrium solutions. Usually, in the majority of tests or applications, the boundary conditions are conceived with the aim of either reproducing the experimental gripping  conditions in the tests  or reflecting the loading conditions of theoretical analysis. These external solicitations are generally  imposed by monotonically  increasing/decreasing loads/displacements in specific positions of the specimen with the purpose of obtaining a particular stress, strains or displacement field. However, due to the onset of failure processes and fracture propagation, the linearity of the solution is compromised and the redistribution of the stress field may lead to simultaneous reduction in load and displacement or, in other words, featuring  snap-back behaviours. This fact jeopardizes the convergence of the simulation employing load or displacement controlled boundary conditions. Although there are methods that consider changes in the direction of the load-displacement curves, for instance the Riks method \cite{RIKS1979}, other techniques have been developed in order to overcome  this kind of instabilities and therefore solving these issues  in an efficient manner.

In this setting, Tvergaard \cite{TVERGAARD1976} proposed an alternative to capture fluctuations in the load-displacement curves by finding a variable that increases monotonically during the simulation. In this way, the control is applied in such variable and the loads and displacements at the boundary are computed  as output variables of the finite element analysis. This approach allows the Newton–Raphson algorithm to be used without any further modifications, and has been tested in simulations including sphere fracture in composites made up of random distribution of elastic spheres within an elasto-plastic matrix \cite{SEGURADO2004}, or the investigation of gradient-enhanced dislocation hardening on the mechanics of notch-induced failure \cite{MARTINEZPANEDA17}. Mainly, the control is applied: (i) to the sum of the opening displacements of some nodes ahead of the crack tip, in presence of a unique interface, or (ii) the relative opening within the interfaces along the loading direction, if more than one interface is involved. In this study, we are focused on the former approach, which is concisely described in the following paragraphs with focus on MMB  specimens. 

\begin{figure}[h!]
	\centering
	\includegraphics[width =0.7 \textwidth]{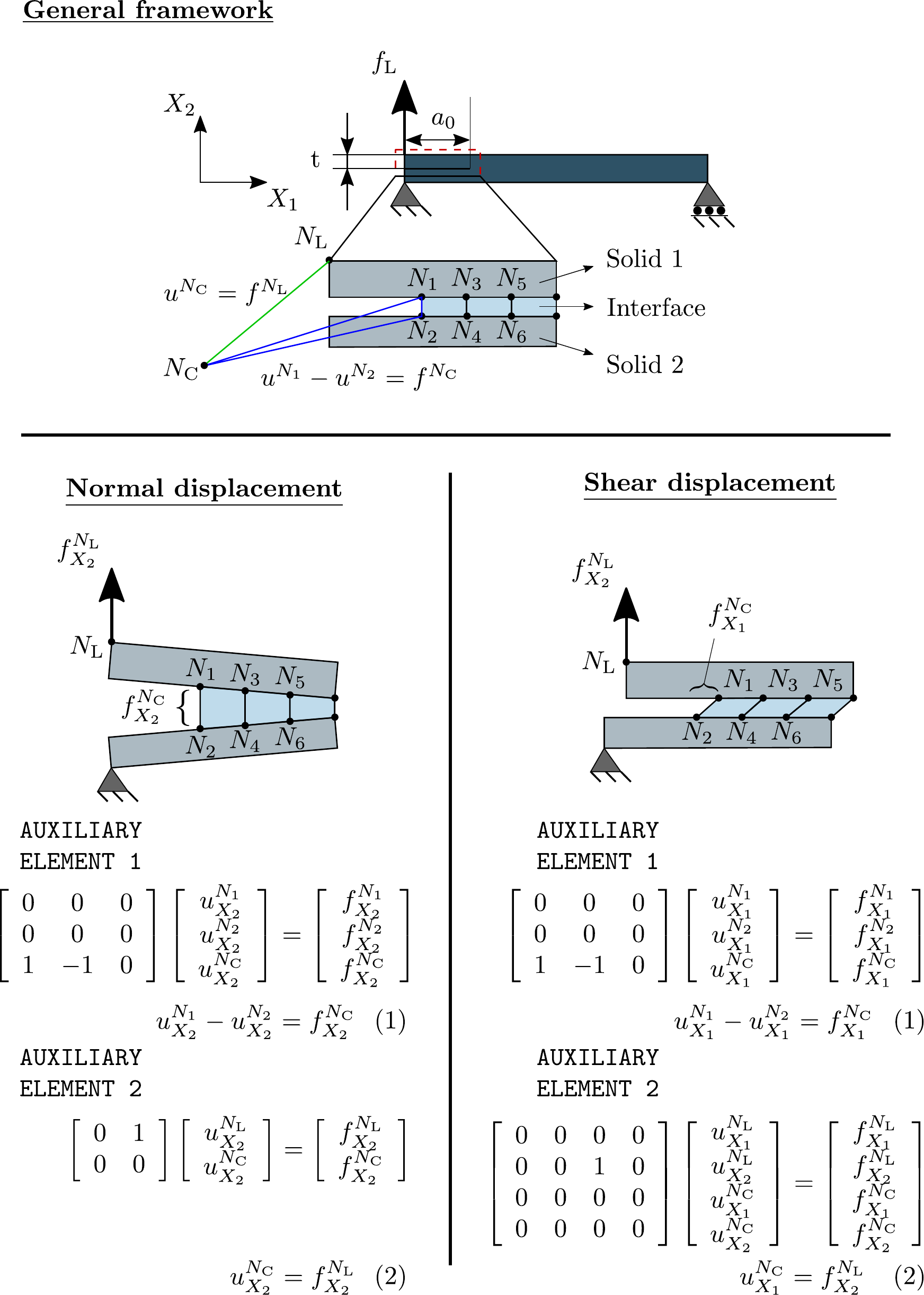}
	\caption{Scheme of the control algorithm: $N_1$-$N_6$ are interface nodes, $N_{\text{C}}$ is the control node and $N_{\text{L}}$ is the node where the boundary conditions are established. $u$ and $f$ stand for nodal displacements and nodal forces at the corresponding nodes.}
	\label{Control_alg}
\end{figure}

To commence the description of this control algorithm,  let $N_1$ and $N_2$ be the nodes belonging to upper and bottom surfaces of the interface, respectively, and $N_{\text{C}}$ a dummy node that can be placed at any point, as depicted in Fig.~\ref{Control_alg}. Likewise, $N_{\text{C}}$ will be the control node and $N_{\text{L}}$ the node in which the load or displacement conditions at the boundary are applied. The relative displacement at the interface, corresponding to the global basis $\lbrace X_1$, $X_2 \rbrace$, can be related to the control node $N_{\text{C}}$ by making use of the \texttt{AUXILIARY ELEMENT}s of a FEM code (\texttt{ABAQUS} being used in the  present investigation) as follows


\begin{equation}
\label{control}
\left[ \begin{array}{ccc}  
0 & 0 & 0 \\ 
0 & 0 & 0 \\
1 & -1 & 0 \\
\end{array} \right]  \left[ \begin{array}{ccc} 
u_{X_1}^{N_1} \\
u_{X_1}^{N_2} \\
u_{X_1}^{N_{\text{C}}} \\
\end{array} \right] = \left[ \begin{array}{ccc} 
f_{X_1}^{N_1} \\
f_{X_1}^{N_2} \\
f_{X_1}^{N_{\text{C}}} \\
\end{array} \right],
\end{equation}

\begin{equation}
\label{control2}
\left[ \begin{array}{ccc}  
0 & 0 & 0 \\ 
0 & 0 & 0 \\
1 & -1 & 0 \\
\end{array} \right]  \left[ \begin{array}{ccc} 
u_{X_2}^{N_1} \\
u_{X_2}^{N_2} \\
u_{X_2}^{N_{\text{C}}} \\
\end{array} \right] = \left[ \begin{array}{ccc} 
f_{X_2}^{N_1} \\
f_{X_2}^{N_2} \\
f_{X_2}^{N_{\text{C}}} \\
\end{array} \right],
\end{equation}
\noindent
where $u_{X_1}$ and $u_{X_2}$ are nodal displacement in global directions and $f_{X_1}$ and $f_{X_2}$ are the corresponding nodal forces. In the FE analysis the relative displacement at the interface, $\left(u_{X_1}^{N_1}-u_{X_1}^{N_2}\right)$ or $\left(u_{X_2}^{N_1}-u_{X_2}^{N_2}\right)$, is set by prescribing the nodal force  of the control node $f_{X_1}^{N_{\text{C}}}$ or $f_{X_2}^{N_{\text{C}}}$. Next,  Eq. \eqref{control} is  employed to prescribed the opening displacement along $X_1$ global direction, whereas Eq. \eqref{control2} is used to set nodal relative displacements along $X_2$ global direction. A flowchart of the  current procedure is given  in Fig.~\ref{Control_alg} for the sake of clarity. A general implementation in a FE package would require the description and adaptation of the method based on the particular characteristics of the FE-code. In any case, a more detailed description can be found in \cite{JUUL2019254}.

Finally, the nodal force at the boundary $f_{X_1}^{N_{\text{L}}}$ or $f_{X_2}^{N_{\text{L}}}$ is equal to the displacement of the control node $u_{X_1}^{N_{\text{C}}}$ or $u_{X_2}^{N_{\text{C}}}$. This relation can be defined through the definition of a new \texttt{AUXILIARY ELEMENT} following

\begin{equation}
\label{control3}
\left[ \begin{array}{cc}  
0 & 1  \\ 
0 & 0 \\
\end{array} \right]  \left[ \begin{array}{ccc} 
u_{X_2}^{N_{\text{L}}} \\
u_{X_2}^{N_{\text{C}}} \\
\end{array} \right] = \left[ \begin{array}{ccc} 
f_{X_2}^{N_{\text{L}}} \\
f_{X_2}^{N_{\text{C}}} \\
\end{array} \right],
\end{equation}

\begin{equation}
	\label{control4}
\left[ \begin{array}{cccc}  
	0 & 0 & 0 & 0 \\ 
	0 & 0 & 1 & 0 \\
	0 & 0 & 0 & 0 \\ 
	0 & 0 & 0 & 0 
\end{array} \right]  \left[ \begin{array}{c} 
	u_{X_1}^{N_{\text{L}}} \\
	u_{X_2}^{N_{\text{L}}} \\
	u_{X_1}^{N_{\text{C}}} \\
	u_{X_2}^{N_{\text{C}}} 
\end{array} \right] = \left[ \begin{array}{c} 
	f_{X_1}^{N_{\text{L}}} \\
	f_{X_2}^{N_{\text{L}}} \\
	f_{X_1}^{N_{\text{C}}} \\
	f_{X_2}^{N_{\text{C}}}
\end{array} \right],
\end{equation}
\noindent
where Eq. \eqref{control3} relates the reaction of the normal opening displacement at the interface $u_{X_2}^{N_{\text{C}}}$ to the nodal force in $X_2$ direction $f_{X_2}^{N_{\text{L}}}$ and Eq. \eqref{control4} associates the reaction of the shear displacement at the interface $u_{X_2}^{N_{\text{C}}}$ to the nodal force in $X_2$ direction $f_{X_2}^{N_{\text{L}}}$.

It should be mentioned that previous operators usually represent a stiffness matrix that relates nodal displacements to nodal forces, so that the components of such matrix should have [Force/displacement] dimensions. Differing from this, the capabilities of the \texttt{AUXILIARY ELEMENT}s are employed in the present analysis in a different way:
\begin{itemize}
	\item The \texttt{AUXILIARY ELEMENT}s, presented in Eqs. \eqref{control}-\eqref{control2},  relate the relative nodal displacement at the interface to the displacements of the control node, so that the components of the matrix associated with this \texttt{AUXILIARY ELEMENT} are dimensionless. Nevertheless, the conventional finite element nomenclature remains for the sake of the consistency. That is, the variable $f^{N_{\text{C}}}$ is calculated as a force unknown in the global system of the Finite Element model, but this variable actually stands for the nodal displacement of the node $N_{\text{C}}$. On the contrary, $u^{N_{\text{C}}}$ represents the reaction force of such node.
	\item The \texttt{AUXILIARY ELEMENT}s, displayed in Eqs. \eqref{control3}-\eqref{control4}, relate the nodal force of the control node $N_{\text{C}}$ to the nodal forces at the boundary, specifically to the node $N_{\text{L}}$ in the current case. Hence, the components of the matrix associated with these \texttt{AUXILIARY ELEMENT}s are dimensionless. In this way, the reaction force of the control node, named as $u^{N_{\text{C}}}$, is related to the nodal force at the boundary, represented by the variable $f^{N_{\text{L}}}$. The specific components of the \texttt{AUXILIARY ELEMENT} will determine the relationship between the degrees of freedom and the direction of the nodal forces corresponding to $N_{\text{C}}$ and $N_{\text{L}}$ nodes.
\end{itemize}

Note however that in situations where  more than one pair of nodes are implied in the process, that is, the \texttt{AUXILIARY ELEMENT}s  (Eqs. \ref{control}-\ref{control2}) are applied to additional pair of nodes, the nodal force of the control node $f^{N_{\text{C}}}$ represents the sum of the opening displacement of each paired nodes. This scheme may be useful to track the global tendency at the interface instead of focusing in a particular pair of nodes.

These equations are  added to the global stiffness matrix of the system in order to compute  the unknown variables, in this case the displacement in the control node $u^{N_{\text{C}}}$ and the displacement of the boundary node $u^{N_{\text{L}}}$. It is worth to emphasize that $u^{N_{\text{C}}}$, through the \texttt{AUXILIARY ELEMENT} materialised in the Eqs. \eqref{control}-\eqref{control2}, corresponds to the reaction force of the boundary node: $u^{N_{\text{C}}} = f^{N_{\text{L}}}$. Therefore, both displacement and force at the boundary, $u^{N_{\text{L}}}$ and $f^{N_{\text{L}}}$, respectively, are calculated as any degree of freedom of the system and they may present non-monotonic behaviour.

As a summary, the control algorithm is outlined in the following scheme:

\begin{enumerate}
	\item Define a control node $N_{\text{C}}$ anywhere.
	\item Prescribe the opening displacement along the interface by means of the control node $f^{N_{\text{C}}}$ and the \texttt{AUXILIARY ELEMENT} 1:\\
	$u^{N_1}-u^{N_2} = f^{N_{\text{C}}}$
	\item Relate the displacement of the control node to the nodal force at the boundary through the \texttt{AUXILIARY ELEMENT} 2:\\
	$u^{N_{\text{C}}} = f^{N_{\text{L}}}$
	\item Include these equations or constraints to the global stiffness matrix.
	\item Obtain $u^{N_{\text{L}}}$ and $f^{N_{\text{L}}}$ as part of the system solution.
\end{enumerate}

\section{Application of the geometrically nonlinear LEBIM to general mixed mode fracture tests: DCB, MMB and ENF}
\label{app}
This Section outlines the validation of the proposed geometrically nonlinear LEBIM for its application to general mixed-mode fracture tests. In particular, we specialize this procedure to numerical-experimental correlation of well established  tests: (i) DCB test for fracture Mode I conditions, (ii) MMB test for mixed-mode fracture conditions and (iii) ENF test for fracture Mode II characterization. With the purpose of testing the accuracy of the interface   approach presented above,    predictions of  such tests, involving delamination events under  different loading conditions, are compared with the experimental data extracted from \cite{Reeder90}. Note also that other authors \cite{Turon06,camanhodavila2003} have previously assessed their corresponding interface  decohesion elements with respect to these experiments. It is worth mentioning that, in the following simulations, the in-plane stiffness of the current LEBIM is set equal to zero, neglecting the in-plane deformation effects, with the aim of comparing  the performance of the present formulation with respect to alternative   cohesive elements  \cite{Turon06,camanhodavila2003,Reinoso2014}.

\begin{figure}[h!]
	\centering
	\includegraphics[width =0.75 \textwidth]{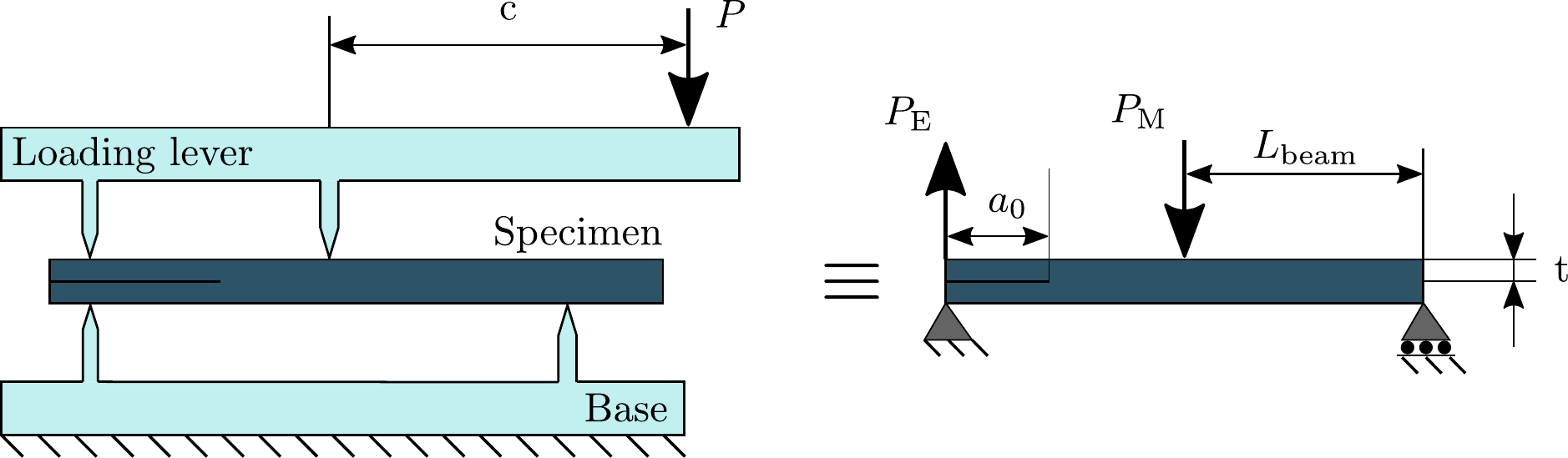}
	\caption{Mixed Mode Bending test: boundary conditions and specimen dimension. $P_{\text{E}}$ and $P_{\text{M}}$ represent the loads applied at the left end and the middle of the specimen. $L_{\text{beam}}$ and t stand for the semi-length and one-arm thick of the coupon, whereas $a_0$ indicates the length of the initial delamination.}
	\label{MMB_test}
\end{figure}

Simulations are  carried out using specimens with the following geometrical dimensions: 50 mm in length of half-span $L_{\text{beam}}$, 25 mm in width and 1.55 mm half-thickness $t$. This set up is defined  according to the test configuration specified in Fig.~\ref{MMB_test}, and it is usually denominated as MMB method that  allows   different mixed-mode fracture ratios  using the same specimen configuration to be assessed. This can be achieved via the definition of a parametric length $c$, whose value can be accordingly set for ranging from pure fracture Mode I to pure fracture  Mode II and covering a wide mixity of ratios. Based on this configuration and recalling standard Bernoulli beam theory, the ratio between the middle and end forces, denoted by    $P_{\text{M}}$ and $P_{\text{E}}$, respectively,  can be related to 
specific mixed-mode ratios $G_{II}/G_T$ (see \cite{camanhodavila2003} for further details).

Additionally to the previous characteristics, an initial crack length $a_0$ is defined  between 30 mm and 40 mm in order to achieve a stable crack propagation, Fig.~\ref{MMB_test}. The corresponding pre-crack lengths $a_0$, the length of the lever $c$ and the  relation between the end and the middle load $P_{\text{M}}/P_{\text{E}}$  for each of the configurations analysed herein are detailed  in Table \ref{a0}, emphasizing the mixed-mode fracture ratios.  The material properties of the laminates (AS4/PEEK composite) are reported in Table \ref{AS4}. The specimens are  composed of 24 unidirectional plies, employing a Kapton film to induce the initial crack length.

\begin{table}[h!]
	\caption{ Initial crack length $a_0$, length of the lever $c$ and middle-end load ratio  $P_{\text{M}}/P_{\text{E}}$, according to Fig.~\ref{MMB_test}, for different mixed-mode ratios $G_{II}/G_{T}$.}
	\centering
	\begin{tabular}{cccccc}
		\toprule
		$G_{II}/G_{T}$   & 0.0 (DCB) & 0.2   &  0.5  &  0.8  &  1.0 (ENF) \\ \midrule
		$a_0$ [mm]       & 32.9     & 33.7  &  34.1  &  31.4 &  39.3 \\
		$c$ [mm]         & -        & 97.4  &  42.2  &  27.6 &  - \\
		$P_{\text{M}}/P_{\text{E}}$ [-] & 0.0   &  1.46  &  2.14 &  2.79 &  $\infty$ \\
		\bottomrule
	\end{tabular}
	\label{a0}
\end{table}

\begin{table}[h!]
	\caption{AS4/PEEK properties.}
	\centering
	\begin{tabular}{cccccc}
		\toprule
		$E_{11}$ & $E_{22}$=$E_{33}$ & $G_{12}$=$G_{13}$ & $G_{23}$ & $\nu_{12}$=$\nu_{13}$ & $\nu_{23}$ \\ \midrule
		129 GPa & 10.1 GPa  & 5.5 GPa & 3.7 GPa & 0.25 & 0.45 \\
		\bottomrule
	\end{tabular}
	\label{AS4}
\end{table}

Regarding the  characteristics of the numerical models, the  adherents are simulated complying with a linear elastic composite material  law, whereas the proposed  LEBIM  is  employed to describe the interface behaviour between the two arms. The baseline numerical model is generated using  approximately 5800 2D plane strain elements for the  discretization of the entire model, where  about  300 of those elements  correspond to interface elements that were equipped with LEBIM.  The undeformed mesh size at the region of interest is  around 0.25 mm in  width and 0.05 mm in height. Table \ref{Cohesive} shows the input properties for the interface  elements \cite{camanhodavila2003}.

\begin{figure}[h!]
	\centering
	\includegraphics[width =0.7 \textwidth]{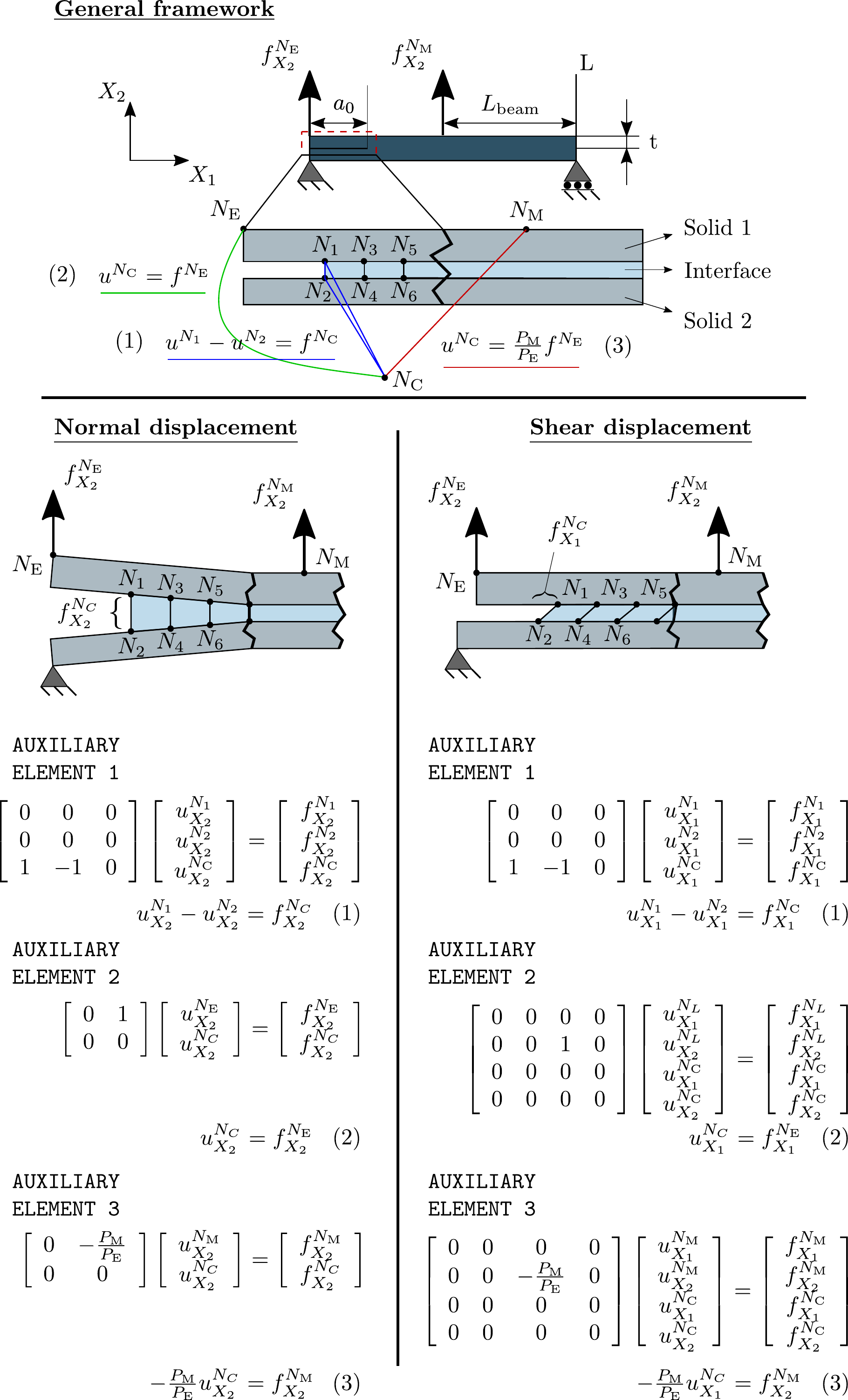}
	\caption{Scheme of the control algorithm for Mixed Mode Bending test: $N_1$-$N_6$ are interface nodes, $N_{\text{C}}$ is the control node and $N_{\text{L}}$ is the node where the boundary conditions are established. $u$ and $f$ stand for nodal displacements and nodal forces in the corresponding nodes. $N_{\text{E}}$ and $N_{\text{M}}$ correspond to the nodes located  at the left end and at the middle, respectively, at the top surface of the upper adherent. }
	\label{Control_MMB}
\end{figure}

The control algorithm described in Section \ref{snapback} is also applied in the current  simulations in order to preclude numerical difficulties for the achievement of converged equilibrium solutions. Such procedure is correspondingly    adapted to each configuration (DCB, MMB and ENF) according to the mixed mode of the tests and the relative displacement at the crack tip. Additionally, a new \texttt{AUXILIARY ELEMENT} is  defined in order to establish the relationship between the forces at the end and at the middle of the specimen, $f^{N_{\text{E}}}_{X_2}$ and $f^{N_{\text{M}}}_{X_2}$ respectively (Section \ref{snapback}). Note the distinction between beam-theory values $P_\text{M}$ and $P_\text{E}$ and the finite element values $f^{N_{\text{E}}}_{X_2}$ and $f^{N_{\text{M}}}_{X_2}$ to denote the nodal boundary forces. In this way, the relation $P_\text{M}/P_\text{E}$ is constant during the test according to the length of the lever $c$ \cite{camanhodavila2003}, as shown in Table \ref{a0}, and such relation is imposed in the FE simulation. Fig.~\ref{Control_MMB} comprehensively details  the procedure to apply normal or shear separation at the crack tip and the way through which  the link between  the nodal force at the boundary $f^{N_{\text{E}}}_{X_2}$ and $f^{N_{\text{M}}}_{X_2}$ with respect to the control node is  constructed in order to obtain  the desired relation.

The normal displacement depicted in  Fig.~\ref{Control_MMB} is  applied in the DCB  and MMB tests. In this regard, with respect to the MMB simulations, we set  the ratios  $G_{II}/G_{T} = 0.2$ and $G_{II}/G_{T} = 0.5$ using  this normal displacement.  Moreover,  in additional computations, the shear displacement Fig.~\ref{Control_MMB} is  employed for the simulation of MMB configurations  with $G_{II}/G_{T} = 0.8$ and for the ENF test. 

\begin{figure}[h!]
	\centering
	\includegraphics[width =0.7 \textwidth]{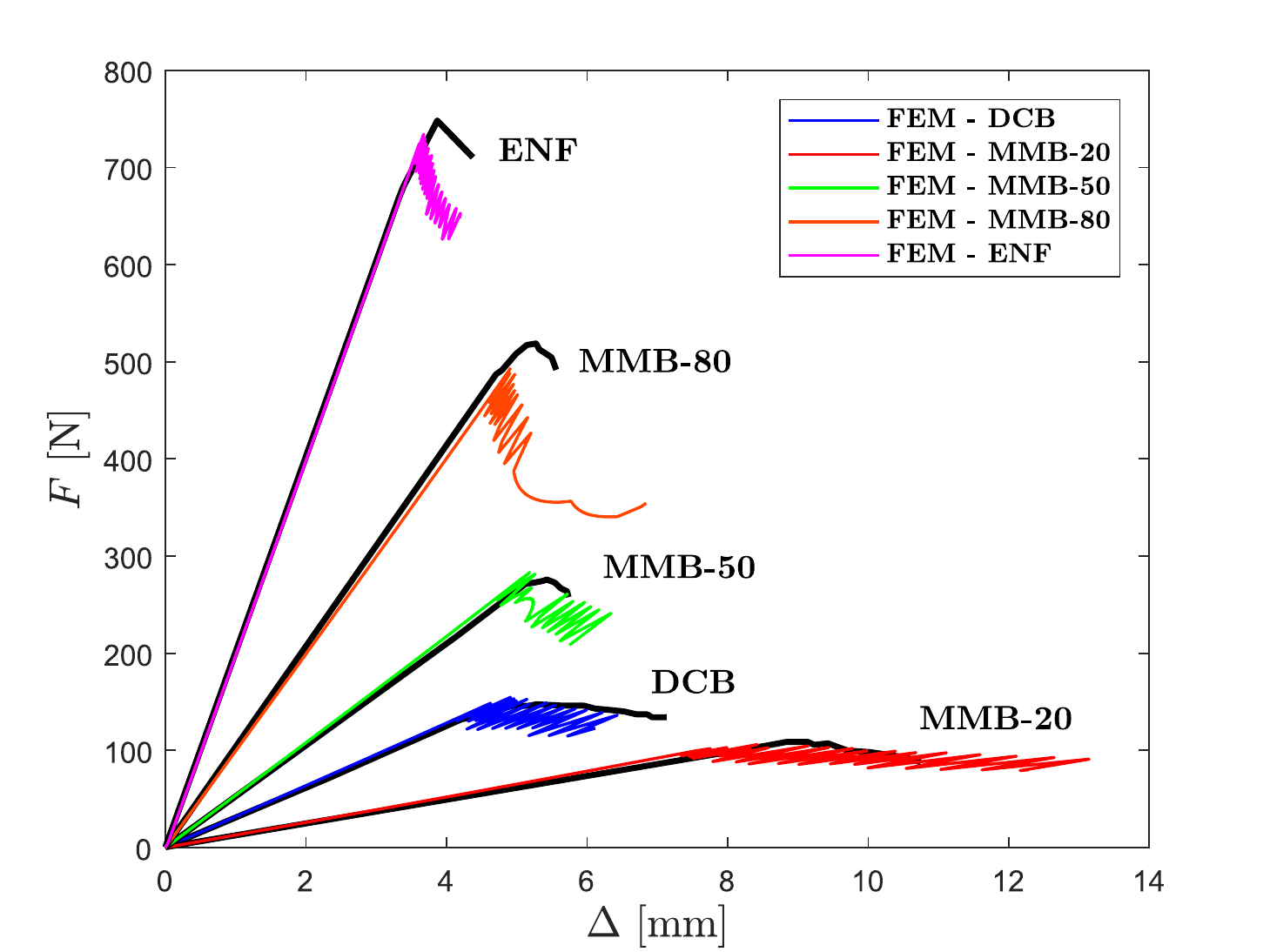}
	\caption{Correlation between experimental \cite{Reeder90} and simulated tests corresponding to DCB, MMB and ENF experiments, including mixed mode ratios $G_{II}/G_{T} = $ [0.0 0.2 0.5 0.8 1.0].}
	\label{Correlation}
\end{figure}

Fig.~\ref{Correlation} shows the correlation between experimental and numerical results concerning load-displacement curves. In this graph, noticeable snap-back effects throughout  the crack propagation in numerical simulations can be identified. This is associated with the boundary conditions imposed in the analysis, in this case an increasing separation (normal or shear) between the crack flanks. Note also that despite the fact  that current  boundary conditions do not exactly replicate the experimental gripping conditions, numerical predictions are in very close agreement with respect to the tests data, in both the linear elastic and crack progression regions of the evolutions.

\begin{table}[h!]
	\caption{Linear Elastic Brittle Interface Model properties \cite{camanhodavila2003}.}
	\centering
	\begin{tabular}{cccccc}
		\toprule
		$t_{\text{n}}^c$ [MPa] & $t_{\text{s}}^c$ [MPa] & $G_{Ic}$ [kJ/m$^2$] & $G_{IIc}$ [kJ/m$^2$] & $\eta$ [-]  \\ \midrule
		80 & 100  & 0.969 & 1.719 & 2.284  \\
		\bottomrule
	\end{tabular}
	\label{Cohesive}
\end{table}

Moreover, interestingly, the evaluation of the mixed mode ratio $G_{II}/G_{T}$ can be performed in a  straightforward manner using LEBIM, due to only the tractions at the crack tip are required as 

\begin{equation}
\label{mixed_mode_LEBIM}
G_{II}/G_{T} = \dfrac{t_{\text{s}} ^2/(2 k_{\text{s}})}{\left< t_{\text{n}} \right>_+^2/(2 k_{\text{n}}) + t_{\text{s}} ^2/(2 k_{\text{s}})} = \dfrac{1}{ \left(\dfrac{\left< t_{\text{n}} \right>_+}{t_{\text{s}}} \right)^2 \dfrac{k_{\text{s}}}{k_{\text{n}}} + 1}.
\end{equation}

\begin{figure}[h!]
	\centering
	\includegraphics[width =0.6 \textwidth]{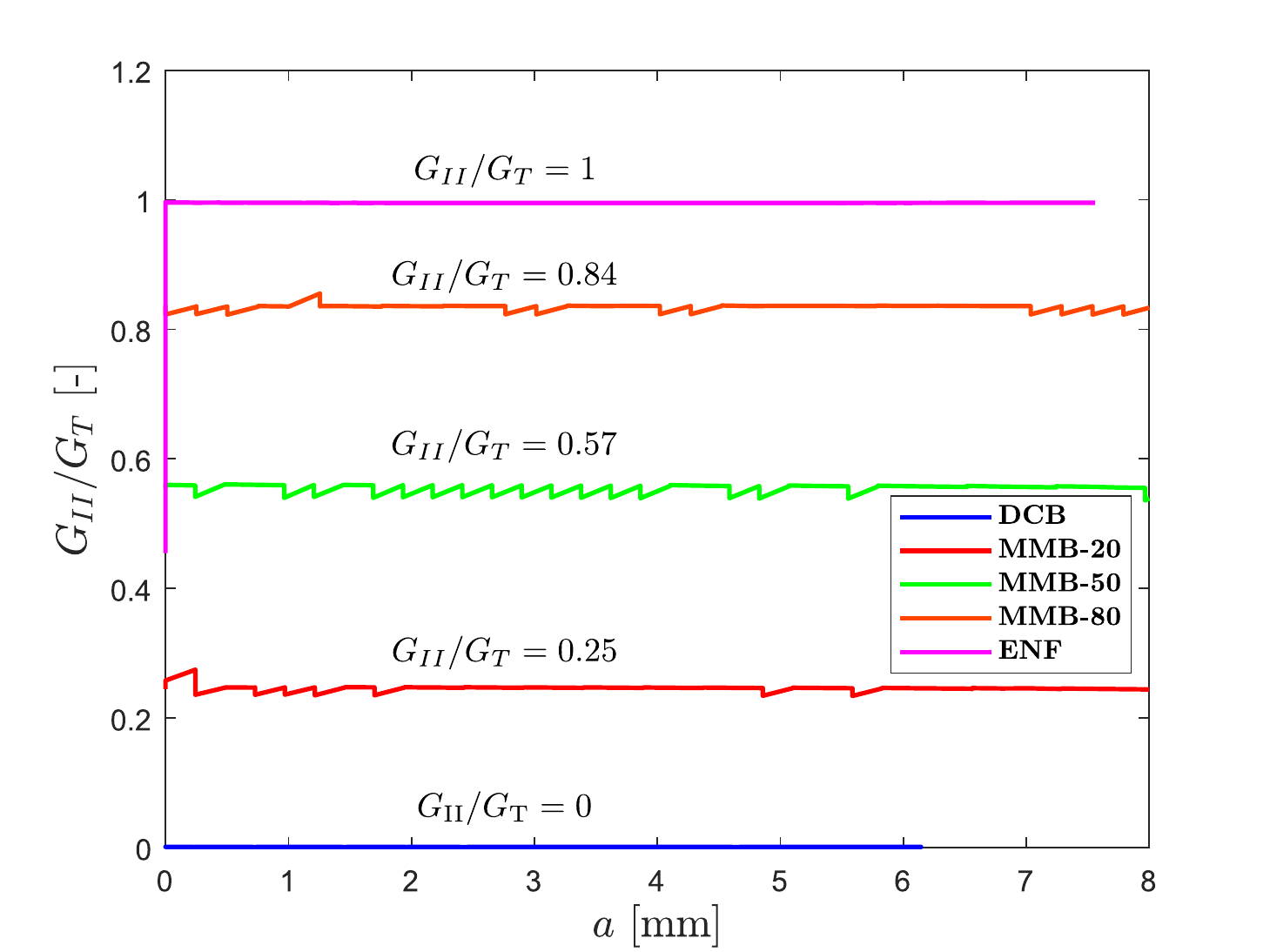}
	\caption{Mixed mode evolution $G_{II}/G_{T}$ at the crack tip (Eq. \eqref{mixed_mode_LEBIM}) corresponding to computational models of DCB, MMB and ENF, including beam-theory mixed mode ratios $G_{II}/G_{T} = $ [0.0 0.2 0.5 0.8 1.0] \cite{camanhodavila2003}.}
	\label{GII_GT}
\end{figure}

Fig.~\ref{GII_GT} depicts the evolution of the mixed mode $G_{II}/G_{T}$ for each configuration according to the expression given in Eq. \eqref{mixed_mode_LEBIM} as a function of the crack length. Such curves present a constant value during the crack growth in conjunction  with some fluctuations lower than the 10\% with respect to their  mean values. Table \ref{Results_MMB} reports  the qualitative comparison between the  experimental and numerical results, where the maximum force $F_{\text{max}}$ and the mixed mode ratio $G_{II}/G_{T}$ are detailed. In the numerical column, the mean value of the curves in Fig.~\ref{GII_GT} is provided.

\begin{table}[h!]
	\caption{Experimental \cite{Reeder90} versus numerical results. $F_{\text{max}}$ is the maximum load in the test, experimental $G_{II}/G_{T}$ is that reported in \cite{Reeder90} (based on beam theory) and Eq. \ref{mixed_mode_LEBIM} at the crack tip is employed in numerical $G_{II}/G_{T}$.}
	\centering
	\begin{tabular}{|cc|cc|}
		\toprule
		\multicolumn{2}{|c|}{Experimental} & \multicolumn{2}{|c|}{Numerical} \\ \hline \addlinespace[1ex] 
		 $G_{II}/G_{T}$ & $F_{\text{max}}$ [N] & $G_{II}/G_{T}$ & $F_{\text{max}}$ [N]  \\ \midrule
		0.00 & 147.5 & 0.00 & 154.6  \\
		0.20 & 108.7 & 0.25 & 106.0  \\
		0.50 & 275.8 & 0.57 & 283.1  \\
		0.80 & 518.7 & 0.84 & 492.8  \\
		1.00 & 748.0 & 1.00 & 734.0  \\
		\bottomrule
	\end{tabular}
	\label{Results_MMB}
\end{table}

Based on the current results, it is possible to state that the current  formulation combining continuum elements and Traction Separation Laws relying on the LEBIM enables capturing  the initiation and evolution of delamination events  under Mode I, Mode II and mixed mode fracture conditions. Furthermore, from a computational perspective, in view of the Fig.~\ref{Correlation}, the control algorithm produces the characteristic snap-back curves during the crack propagation that do not appear in experimental data. This   stems from differences between the numerical and the experimental loading conditions.  Thus, whereas  the experimental gripping system  did  not allowed backward movements whereby the displacements or forces are applied,  numerical simulations prescribed an opening displacement at the crack tip in pursuit of the computational convergence, even though enabling the  snap-back behaviour. Note however that, in spite of such discrepancies in terms of the supporting conditions between the experimental and numerical data, the maximum loads for each configuration are  in very satisfactory agreement  (less than 5\% error in the worst case scenario), revealing the accuracy of the proposed LEBIM. Regarding the mixity of the MMB tests, the mixed mode value derived from LEBIM formulation (Eq. \ref{mixed_mode_LEBIM})  slightly differs from the   predicted value of $G_{II}/G_{T}$ using  the classical Bernoulli beam theory. This small deviation could be attributed to the fact that the current form of such classical theory does not account for geometrically nonlinear effects that are especially relevant for the MMB configurations. 

\section{Application of the geometrically nonlinear LEBIM for  DCB specimens with hierarchical trapezoidal interfaces}
\label{Trap}

\subsection{LEBIM validation by means of experimental-numerical correlation of structured interfaces in DCB tests}
\label{Trap_validation}
This Section addresses the applicability of the proposed geometrically nonlinear LEBIM for the analysis of novel interface profile using structured patterns. This is within the scope of the research activities previously carried out by the authors in \cite{GarciaGuzman19,garcia-guzman} in which additive layer manufacturing (ALM) capabilities for composite materials have been exploited. 

\begin{figure}[h!]
	\centering
	\includegraphics[width =0.75 \textwidth]{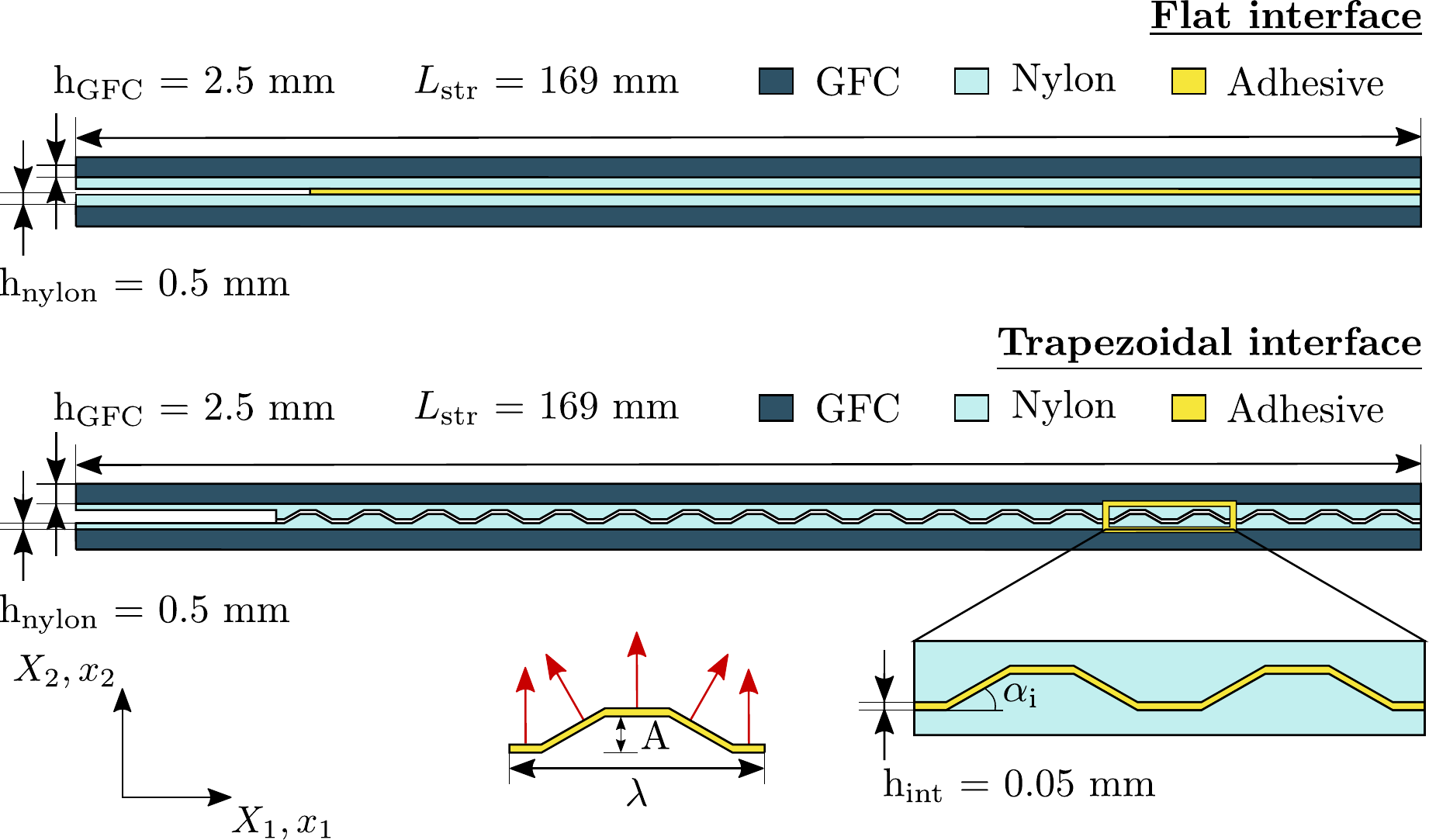}
	\caption{Double Cantilever Beam specimen with flat and trapezoidal interface. Materials: glass-fibre composite (GFC), nylon and adhesive. Dimensions: length $L_{\text{str}}$, height of glass-fibre composite $\text{h}_{\text{GFC}}$, height of nylon in the bulk part $\text{h}_{\text{nylon}}$, amplitude A and wavelength $\lambda$ of the trapezoidal interface.}
	\label{specimen}
\end{figure}

In particular, we specialize the manufacturing of trapezoidal interface DCB specimens using Glass Fiber Composite (GFC) and nylon. For validation purposes, the flat specimen and one of the patterned configurations experimentally tested in \cite{garcia-guzman} were analysed using the interface framework in Section \ref{interface}. The overall dimensions of the coupons employed in the FE simulations are:

\begin{itemize}
	\item Flat interface: $\text{h}_{\text{GFC}} = 2.5$ mm, $\text{h}_{\text{nylon}} = 1.5$, $\text{h}_{\text{int}} = 0.05$ mm, $L_{\text{str}} = 169$ mm,  according to the scheme of Fig.~\ref{specimen}.
	\item Trapezoidal interface: $\text{h}_{\text{GFC}} = 2.5$ mm, $\text{h}_{\text{nylon}} = 0.5$, $\text{h}_{\text{int}} = 0.05$ mm, $L_{\text{str}} = 169$ mm, $A = 1.7$  and $\lambda = 8$ mm, according to the scheme of Fig.~\ref{specimen}.
\end{itemize}

As the previous case analysed in Section \ref{app}, adherents are simulated using a linear elastic behaviour, no damage emerging then in this part of the specimen, and the adhesive layer is represented by a linear elastic brittle law. Current LEBIM properties along the interface are listed in Table \ref{Tab_LEBIM_0} and the GFC and nylon properties are specified in Table \ref{Tab_fibre}. In the LEBIM, the stiffness relationship is set to $k_{\text{s}}/k_{\text{n}} = 1$. The methodology presented in Sect. \ref{large_disp} was employed to model the interface in the DCB simulations so as to examine the role of the in-plane deformations within the adhesive/interface.

\begin{table}[h!]
	\centering
	\begin{tabular}{|c|c|c|c|c|}
		\hline
		Material & $t_{\text{n}}^c$ [MPa] & $t_{s}^c$ [MPa] & $G_{Ic}$ [J/m$^{2}$] & $G_{IIc}$ [J/m$^{2}$]   \\ \hline
		Adhesive  & 4.0 & 16.0 & 136.3  & 2180  \\ \hline
	\end{tabular}
	\caption{Properties of the adhesive modelled as a LEBIM in the experimental-numerical correlation.}
	\label{Tab_LEBIM_0}
\end{table}

\begin{table}[h!]
	\centering
	\begin{tabular}{|c|c|c|c|c|c|c|c|c|c|c|}
		\hline
		Material  & $E_{11}$ [MPa] & $E_{22}$ [MPa] & $E_{33}$ [MPa] & $\upsilon_{12}$ [-] & $\upsilon_{13}$ [-] & $\upsilon_{23}$ [-] & $G_{12}$ [MPa]  \\ \hline
		GFC    & 25863  & 1221  & 1221 & 0.45 & 0.45 & 0.45 & 778 \\ \hline
		Nylon         & 384    & 384   & 384  & 0.39 & 0.39 & 0.39 & - \\ \hline
	\end{tabular}
	\caption{Properties of the glass-fibre composite (GFC) and nylon.}
	\label{Tab_fibre}
\end{table}

Regarding the FE model, 4-node plane-strain elements (type CPE4 in \texttt{ABAQUS}{\textsuperscript{\textregistered}} library) are employed in the adherents as well as in the interface. In the flat case, around 22k elements constitute the adherents and 576 elements (0.25 mm in length) form the interface region. Conversely, around 140k elements made up the adherents and around 150 elements (0.06 mm in length) form each trapezium.

\begin{figure}[h!]
	\centering
	\subfloat[Flat configuration]{\includegraphics[width=0.49\textwidth]{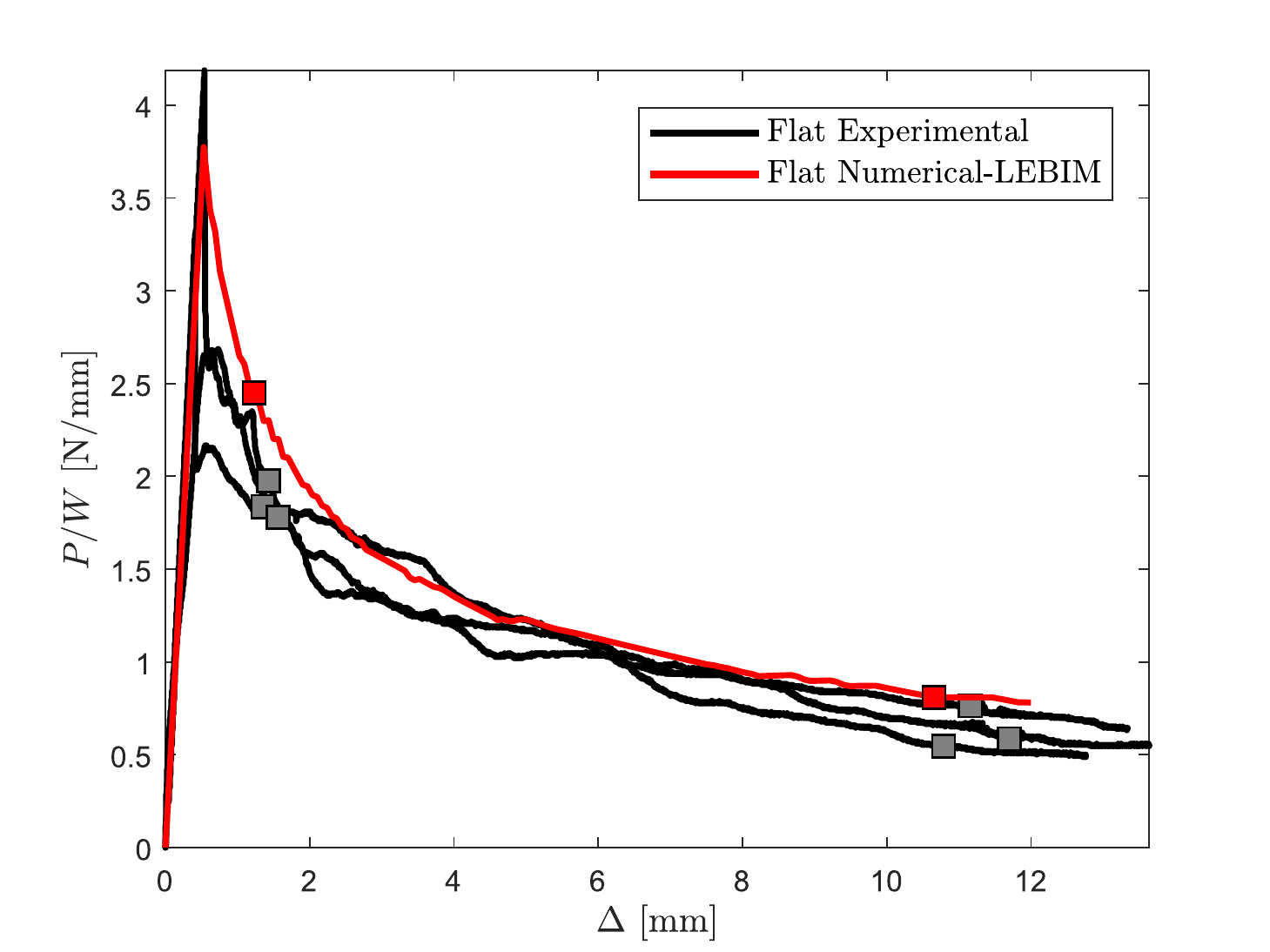}}
	\subfloat[Trapezoidal configuration]{\includegraphics[width=0.49\textwidth]{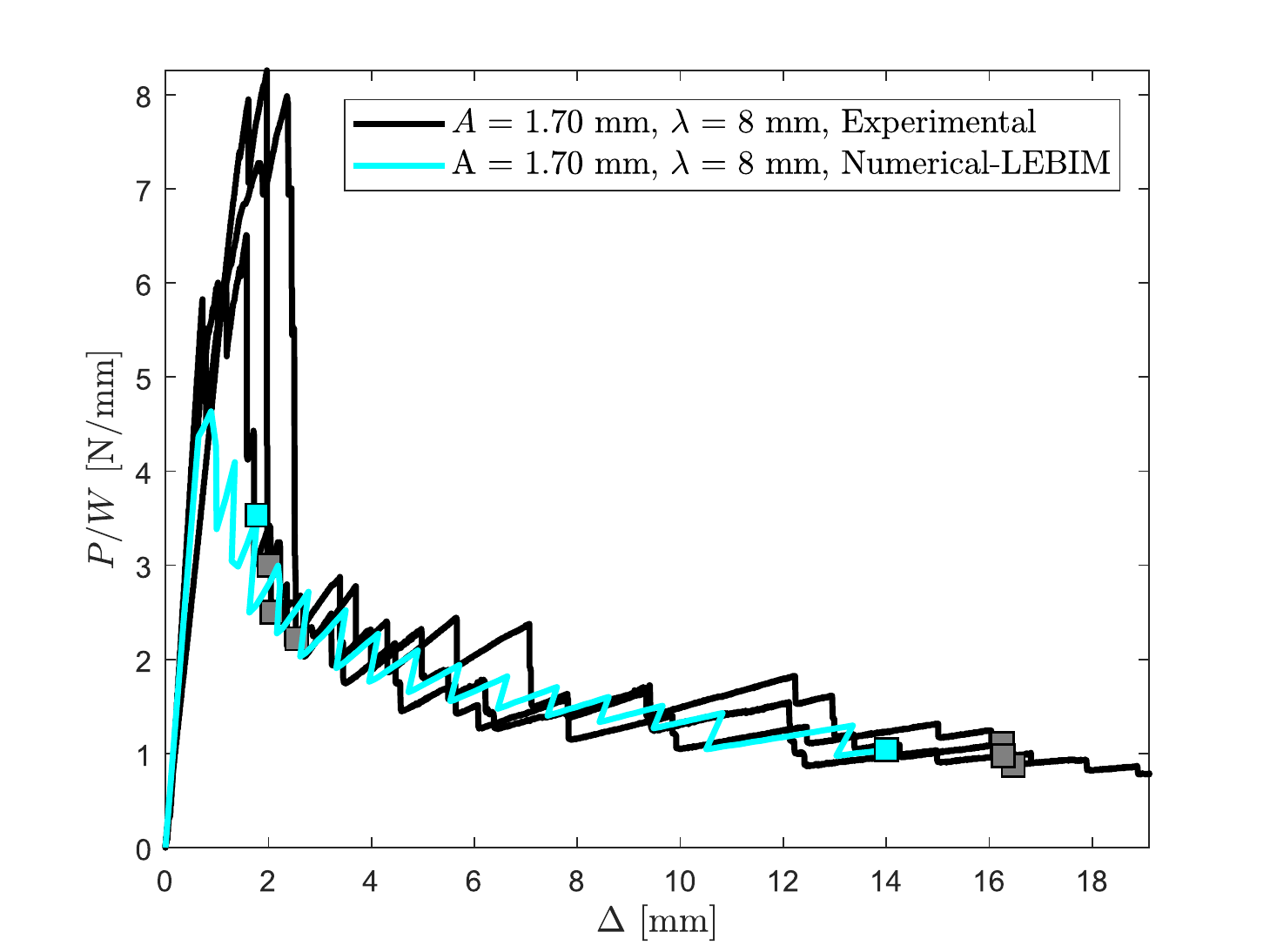}}
	\caption{Experimental-numerical correlation of the load-displacement curves. Square markers represent the points of the curves where the effective crack length reaches $a_{X1} = 10$ mm and $a_{X2} = 70$ mm.}
	\label{Correlation_LEBIM}
\end{figure}

\begin{figure}[h!]
	\centering
	\subfloat[Mixed mode evolution]{\includegraphics[width=0.49\textwidth]{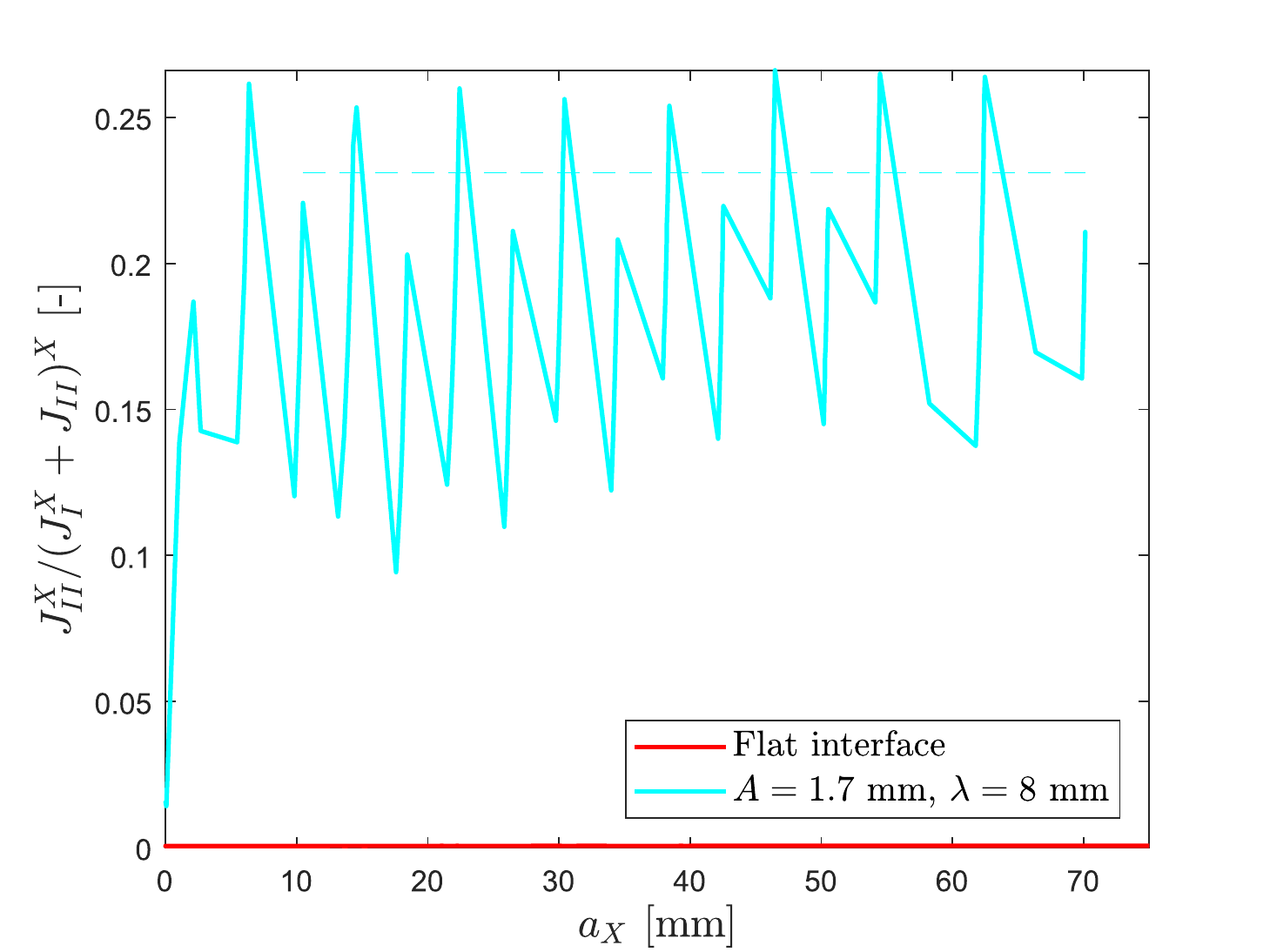}}
	\subfloat[Effective energy release rate evolution evolution]{\includegraphics[width=0.49\textwidth]{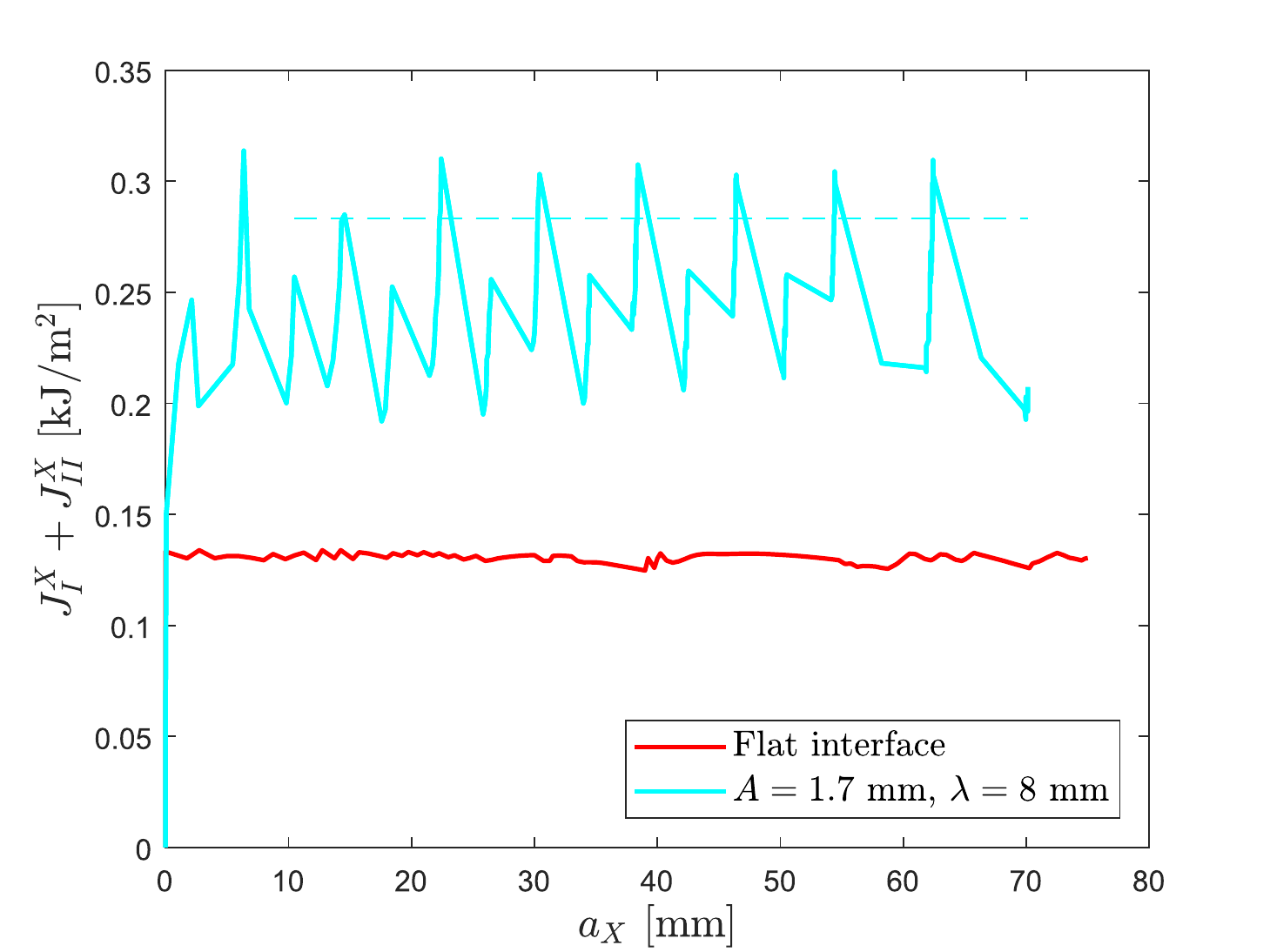}}
	\caption{(a) Numerical evolution of the mixed mode $J_{II}^X/J_{T}^X$ and (b) effective energy release rate $J^{X}$ with respect to the effective crack length $a_X$ in flat and trapezoidal configurations. Dashed lines represent the average value according to the maximum values or peaks of the curves.}
	\label{MM_Jx}
\end{figure}

Two different methods were used to evaluate the fracture energy in the patterned interfaces:

\begin{itemize}
	\item First, the critical energy release rate $G_c$, based on the  standards outlined in \cite{airbus2006} and employed previously in \cite{garcia-guzman}, is determined as the area under the load-displacement curve with respect to the effective or apparent cracked surface (crack length $a_X$ in a 2D analysis) between two different crack lengths, as depicted in Fig.~\ref{Correlation_LEBIM} and Fig.~\ref{F_D}. $a_{X1}$ and $a_{X2}$ included in the same plot were employed in the fracture characterisation according to the expression:
	
	\begin{equation}
	\label{G_LD}
	G_{c}^{\text{LD}} = \frac{A^{\text{LD}}}{a_{X2} - a_{X1}}.
	\end{equation}
	
	\item Second, the effective J-Integral developed in \cite{GarciaGuzman19}, defined as the variation of the potential energy with respect to the horizontal projection of the crack advance ($X_1$ global axis)
	
	\begin{equation}
	\label{J-Integral-Eq}
	J^X = -\frac{d \Pi}{d a_X} = - \int_{\partial \Gamma} \left( \frac{\omega(X_1,X_2)}{cos(\alpha)} \textrm{d}X_2- \frac{t_i}{cos(\alpha)} \frac{\partial u_i}{\partial X_1} \textrm{d}s\right).
	\end{equation}
	
	Dividing the last expression into symmetrical and anti-symmetrical counterparts, the fracture energy developed in Mode I and Mode II can be obtained by means of the tractions and displacements within the interface as
	
	\begin{equation} \label{J-Int-Coh-1-ax}
	J_{I}^{X} = \sum_{k=1}^{n} J_{I, \Gamma_{k}+\Gamma_{k'}}(a_X) = \sum_{k=1}^{n} \int_{\Gamma_{k}} \frac{t_{\text{n}}}{\cos \alpha} \frac{\partial \delta_{\text{n}}}{\partial X_1}\textrm{d}X_1,
	\end{equation}
	
	\begin{equation} \label{J-Int-Coh-2-ax}
	J_{II}^{X} = \sum_{k=1}^{n} J_{II, \Gamma_{k}+\Gamma_{k'}}(a_X) = \sum_{k=1}^{n} \int_{\Gamma_{k}} \frac{t_{\text{s}}}{\cos \alpha} \frac{\partial \delta_{\text{s}}^{\text{s}}}{\partial X_1}\textrm{d}X_1,
	\end{equation}
	
	where $\alpha$ is the angle respect to the horizontal plane (see Fig.~\ref{specimen}) and it depends on the position along the crack path: $\alpha= \alpha(X_1)$. $\Gamma_{k}$ represents the different sections along the profile. The path selected to perform the J-Integral calculations was the upper and lower surface of the interface, from the crack tip to the point where the normal stress becomes null. Fig.~\ref{MM_Jx} and Fig.\ref{Jx_ax_hier} shows the evolution of the $J^{X} = J_{I}^{X}+J_{II}^{X}$ respect to the effective crack length $a_X$. Additionally, for comparing purposes, a mean value of the J-Integral is provided by means of
	
	\begin{equation} \label{Jx_mean}
	\bar{J}_c^{X}  = \frac{1}{a_{X2}-a_{X1}} \int_{a_{X1}}^{a_{X2}}J^{X} \: \textrm{d}a_X.
	\end{equation}
	
	Last expression allows a direct comparison with $G_c$ to be performed due to the fact that for elastic materials $J = G$.
	
\end{itemize}

Nonetheless, in order to properly exploit the large-displacement procedure pinpointed in Section \ref{large_disp} for non-flat interfaces, a pre-process for the interface zone and a slight modification of the \texttt{UMAT} are required. Specifically, an initial rotation of the deformation matrix $\mathbf{F}$ is performed to obtain the strain field expressed in a coordinate system in accordance with the initial midplane of the interface. It is worth emphasising that in presence of structured interfaces, the direction of the path $\alpha_{\text{i}}$ with respect to the global coordinate system ($X_1$, $X_2$) is a function of the position. In this way, Eq. \ref{Ri} points out the operator that is  required in order to get the appropriate reference system:

\begin{equation}
\label{Fi}
\mathbf{F}_{\text{i}} = \mathbf{R}_{\text{i}}^{T} \mathbf{F} \mathbf{R}_{\text{i}},
\end{equation}
\noindent
where

\begin{equation}
\label{Ri}
\mathbf{R_{\text{i}}} = \left[ \begin{array}{cc}  
\cos(\alpha_{\text{i}}) & -\sin(\alpha_{\text{i}}) \\ 
\sin(\alpha_{\text{i}}) & \cos(\alpha_{\text{i}})
\end{array} \right].
\end{equation}
\noindent
This operation should to be performed at each integration point with its corresponding $\alpha_{\text{i}}$ value. A \texttt{PYTHON} script was developed to get the initial slope for each integration point and it was transferred to the \texttt{UMAT} as a dummy initial state variable by means of the \texttt{SDVINI} user subroutine.

Finally, as the variables of the interface elements were expressed in global coordinates, an additional rotation of the stress tensor and Jacobian matrix is  performed.
It is worth mentioning that the output variables of the \texttt{UMAT} should be expressed in the global Cartesian basis.

\begin{figure}[h!]
	\centering
	\includegraphics[width =0.9 \textwidth]{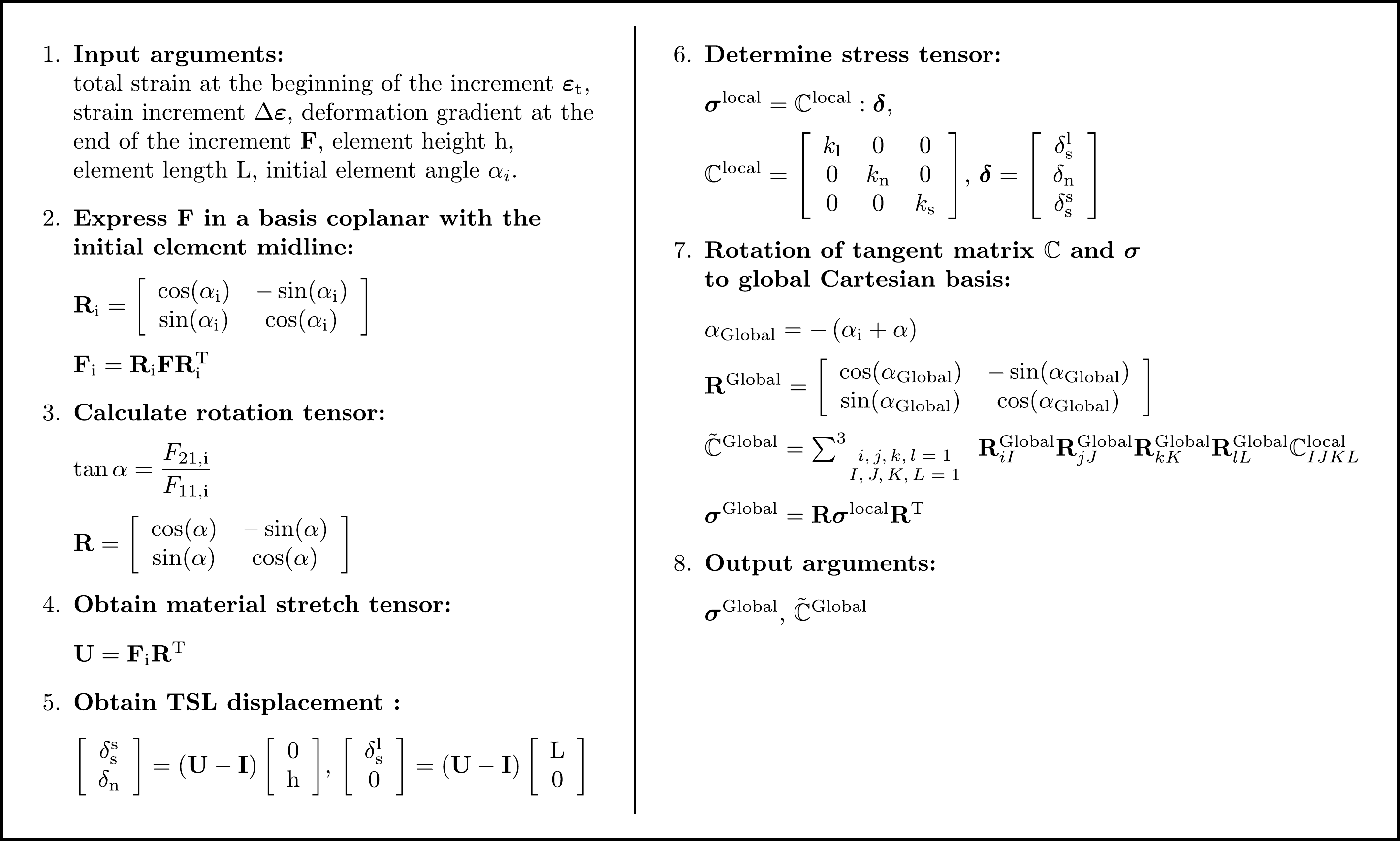}
	\caption{Simplified algorithm for displacement-stress estimation in a Traction Separation Law presenting a curved crack path in interface continuum elements under finite displacement and rotation assumptions.}
	\label{Algorithm_trapz}
\end{figure}

Hence, a rotation of $-(\alpha+\alpha_{\text{i}})$ radians is carried out for achieving equilibrium and getting convergence. Fig.~\ref{Algorithm_trapz} shows the pseudo-code used to calculate the tractions and the Jacobian matrix in curved profiles.

\begin{figure}[h!]
	\centering
	\includegraphics[width =0.95 \textwidth]{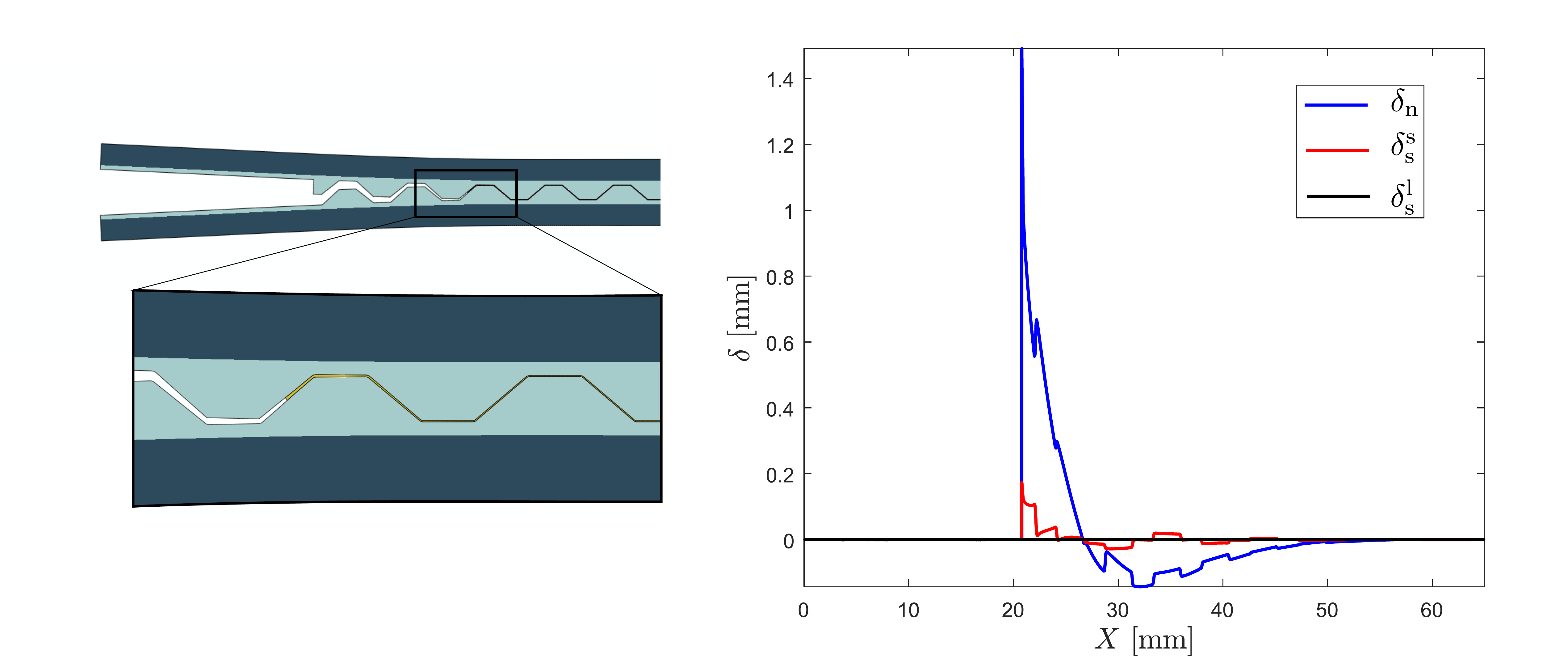}
	\caption{Displacement profile ($\delta_{\text{n}}$, $\delta_{\text{s}}^{\text{s}}$, $\delta_{\text{s}}^{\text{l}}$) along the interface at $\Delta = 2.4$ mm and $P/W = 2.43$ N/mm in the load-displacement curve in Fig.~\ref{Correlation_LEBIM}.b.}
	\label{delta_X}
\end{figure}

The previous algorithm allows the displacement field along the interface to be computed and, consequently, a comparison between normal, shear and in-plane displacements. Fig.~\ref{delta_X} displays a drawing of the deformed DCB specimen and the displacement components ($\delta_{\text{n}}$, $\delta_{\text{s}}^{\text{s}}$, $\delta_{\text{s}}^{\text{l}}$) along the interface length in an intermediate increment of the simulation. It can be observed that $\delta_{\text{n}}$ represents the highest values in the displacement field, followed by the shear displacement $\delta_{\text{s}}^{\text{s}}$. The in-plane deformations can be considered negligible with respect to $\delta_{\text{n}}$ or $\delta_{\text{s}}^{\text{s}}$. In fact, the in plane displacements $\delta_{\text{s}}^{\text{l}}$ do not exceed $1 \times 10 ^{-3}$ mm during the test, that is, the maximum in-plane displacement represent the 0.5\% of the maximum relative shear displacement and the 0.07\% of the maximum relative normal displacement. Hence, neglecting the in-plane deformation in this scenario is an adequate hypothesis (that can be incorporated by setting $k_{\text{l}}=0$).

Fig.~\ref{Correlation_LEBIM} shows the experimental-numerical correlation of the load displacement curves corresponding to the DCB tests in the flat and trapezoidal interfaces. Fig.~\ref{MM_Jx} displays the mixed mode $J_{II}^X/J_{T}^X$, where $J_{T}^X = J_{I}^X + J_{II}^X$, and the $J^X$ evolution with respect to the effective crack length $a_X$ obtained from the FE models.

A good agreement between the curves can be observed in both flat and patterned interfaces. The initial linear-elastic behaviour is captured appropriately as well as the propagation phase, including the unstable crack advance in the trapezoidal case. Furthermore, the prediction of the crack length is in accordance with the experiments. Notwithstanding, the larger discrepancies involving crack length and load-displacement curve occur at the fracture initiation stage. Regarding the mixity of the FE models, the flat configuration led to $B=0$, as expected, whereas the trapezoidal case presents sharp fluctuations along the virtual test whose maximum values are established around $B=0.2$. The J-Integral evolution shares the features of the mixed mode distribution and the average effective energy release rate $\bar{J}_c^X$ is almost twice higher in the patterned scenario than in the flat interface ($\bar{J}_c^X = 243.3$ kJ/m$^2$ in the trapezoidal interface and $\bar{J}_c^X = 130.3$ kJ/m$^2$ in the reference scenario). A summary of the results in the experimental and numerical analysis are included in Table \ref{G_J_exp}.

\begin{table}[h!]
	\centering
	\begin{tabular}{|c|c|c|c|c|}
		\hline
		Configuration & $G_{c}^{\text{LD}}$ [J/m$^{2}$] (Experimental) & $G_{c}^{\text{LD}}$ [J/m$^{2}$] (Numerical) & $\bar{J}_c^{X}$ [J/m$^{2}$] & $\frac{J_{II}}{J_{T}}\rvert_{\text{mean}}$ [-] \\ \hline
		Flat          & 136.3 & 148.7 & 130.3 & 0.0 \\ \hline
		Trapezoidal   & 274.0 & 257.6 & 243.3 & 0.231 \\ \hline
	\end{tabular}
	\caption{Critical energy release rate $G_{c}^{\text{LD}}$ obtained from load-displacement curves (experimental and numerical), mean effective J-Integral $\bar{J}_c^{X}$ and mean mixed mode of the flat and trapezoidal interfaces in the DCB tests. }
	\label{G_J_exp}
\end{table}

\begin{figure}[h!]
	\centering
	\includegraphics[width =0.5 \textwidth]{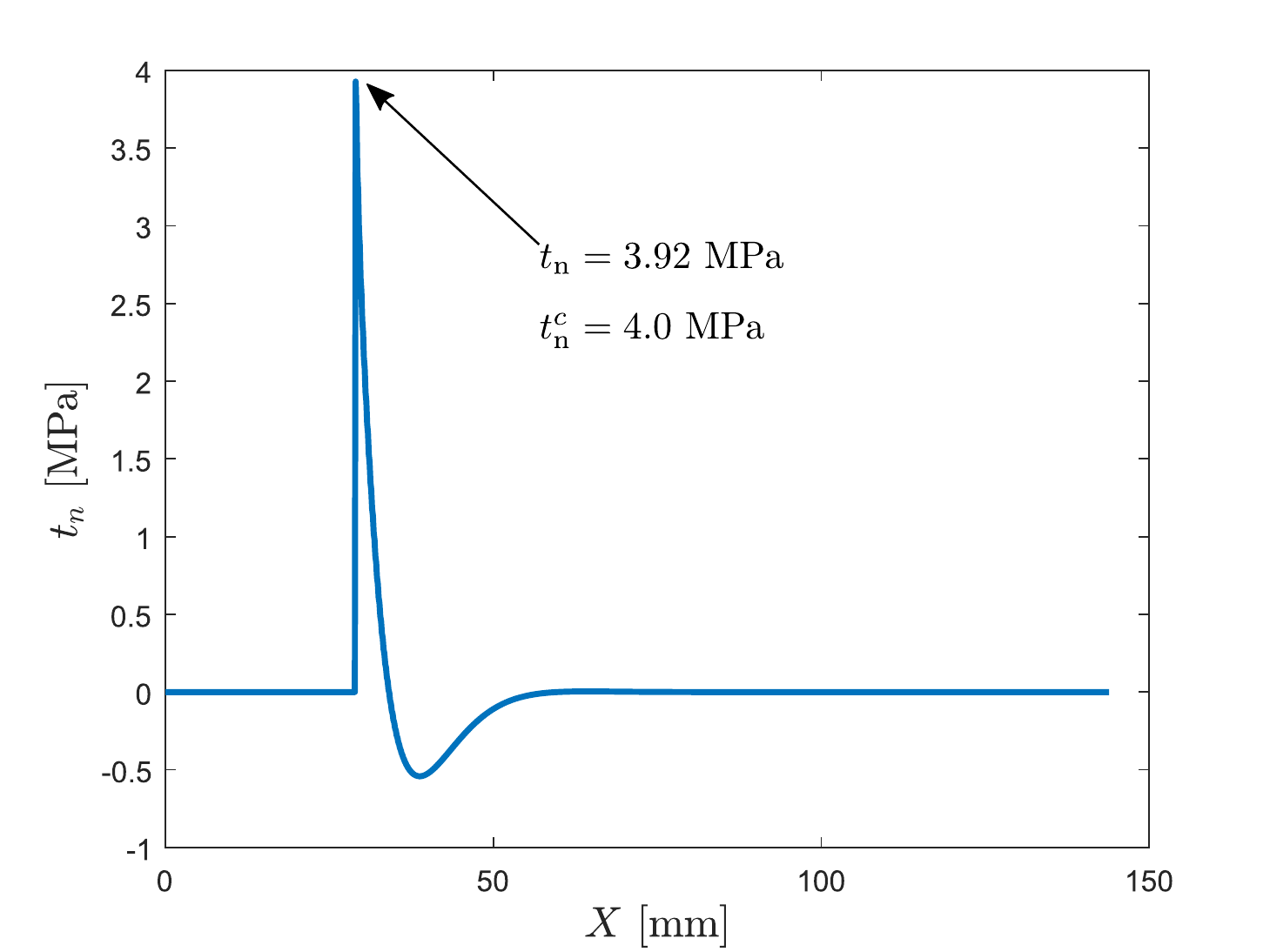}
	\caption{Traction distribution along the interface corresponding to $\Delta= 3.25$ mm and $P/W = 1.50 $ N/mm in the flat DCB test (Fig.~\ref{Correlation_LEBIM}).}
	\label{tn_X}
\end{figure}

In view of the results, it is remarkable that the energy release rate obtained from the area of the load-displacement curve $G_{c}^{\text{LD}}$ in the flat case is higher than the experimental $G_{Ic} = 136.3$ kJ/m$^{2}$. This discrepancy may emerge from the difference between experimental and computational curves at the beginning of the test. Then, if the crack length $a_{X1}$ was selected so that the corresponding point in the load-displacement curve was located at $\Delta \geq 4$ mm, for example, the discrepancy in the fracture toughness will be reduced. On the contrary, the average value of $J^X$ is lower than $G_{Ic}$. This reduced value of the pure Mode I energy is associated with the distribution of stresses along the interface. The value of normal traction $t_{\text{n}}$ just before the crack propagation is lower than the cut-off traction $t_{\text{n}}^c$ established in the TSL, as depicted in Fig.~\ref{tn_X}. Hence, the critical energy release rate calculated from the J-Integral, $\bar{J}_c^{X}$, would be equal to $G_{Ic}$ when the increments of the simulations allow an accurate/perfect traction distribution of the TSL along the interface.

Regarding the patterned configuration, the calculation of the effective fracture toughness is in close agreement: less than 7\% of difference using the load-displacement curves ($G_{c}^{\text{LD}}$) and less than 12\% using the J-Integral approach ($\bar{J}_c^{X}$). Moreover, the mixed mode based on Benzeggah-Kenane criterion is highly accurate for energy predictions: $G_c(B=0.231) = 245.4$ kJ/m$^{2}$, see Eq. \eqref{Gc}.

\subsection{Application: hierarchical trapezoidal interfaces}
\label{Trap_hierarchical}

Double Cantilever Beam tests including a non flat interface between adherents is performed in this Section. This kind of analysis allows the fracture energy to be studied in presence of mixed mode conditions. Furthermore, authors in \cite{garcia-guzman} have demonstrated the significance of the failure modes in epoxy adhesives, which can be modelled assuming brittle behaviour. In this way, LEBIM represents an appropriate tool to describe the crack resistance in these experiments. In addition, the performance of hierarchical crack paths, involving uni-trapezoidal, bi-trapezoidal and tri-trapezoidal patterns, is herewith carried out for comparison purposes.

\begin{figure}[h!]
	\centering
	\includegraphics[width =0.85 \textwidth]{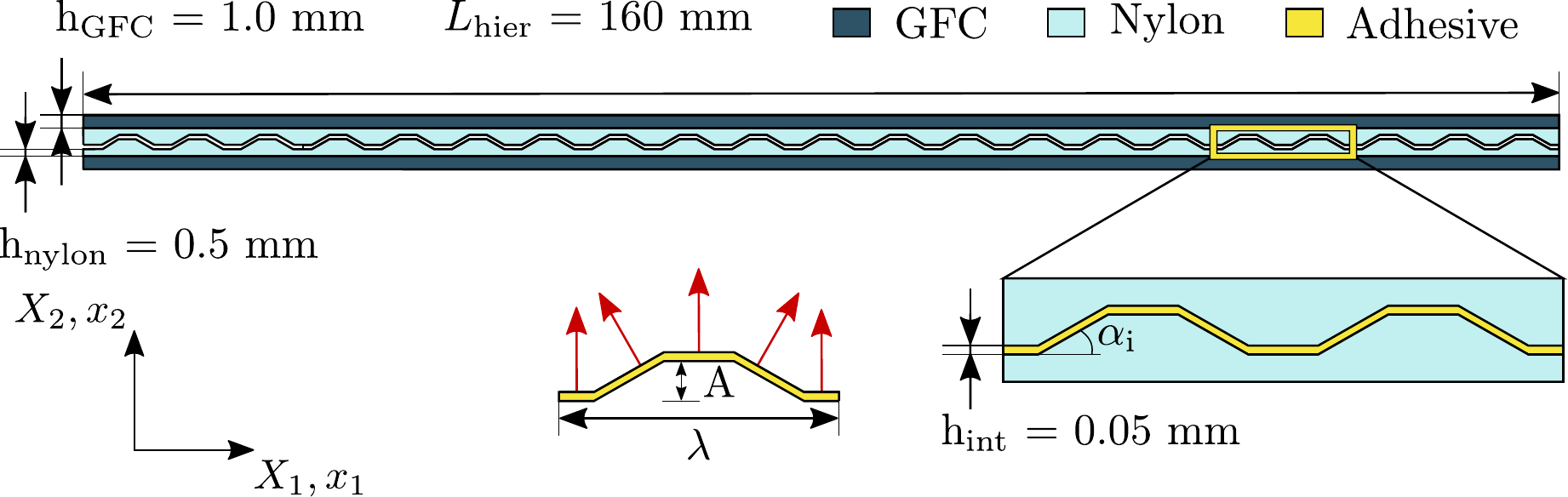}
	\caption{Double Cantilever Beam specimen with a trapezoidal interface. Materials: glass-fibre composite (GFC), nylon and adhesive. Dimensions: length $L_{\text{hier}}$, height of glass-fibre composite $\text{h}_{\text{GFC}}$, height of nylon in the bulk part $\text{h}_{\text{nylon}}$, amplitude A and wavelength $\lambda$ of the trapezoidal interface.}
	\label{specimen_hier}
\end{figure}

3D printed specimens, depicted in Fig.~\ref{specimen_hier}, are 160 mm in length $L_{\text{hier}}$, 20 mm in width W and a total height h of 4.9 mm, where nylon and glass fibre composite (GFC) are used. The bulk part of the coupon consists of 1 mm of GFC and 0.5 mm of nylon, while the layers that form the trapezoidal interface (A = 1.9 mm, $\lambda =$ 8 mm) are made of nylon exclusively. 

\begin{figure}[h!]
	\centering
	\includegraphics[width =0.6 \textwidth]{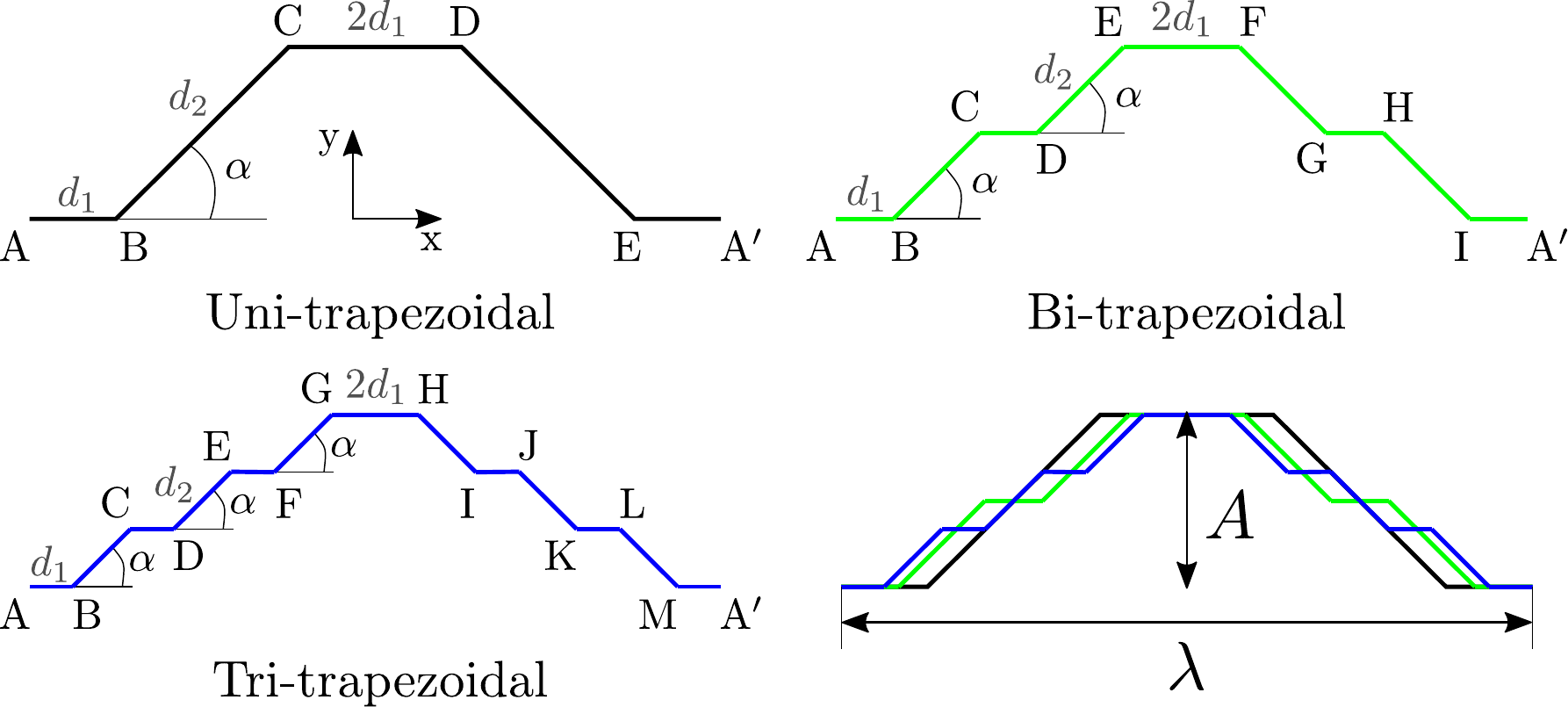}
	\caption{Trapezoidal, bi-trapezoidal and tri-trapezoidal shapes contained in the study. $d_1$ and $d_2$ represent the flat and inclined section length, respectively. $\alpha$ shows the angle in the inclined sections and $A$ and $\lambda$ symbolise the amplitude and the wavelength of each configuration.}
	\label{Hierarchical}
\end{figure}

Regarding the hierarchically-based study, three orders of arrangement are investigated in the sequel: uni-trapezoidal, bi-trapezoidal and tri-trapezoidal patterns. Fig.~\ref{Hierarchical} depicts the geometry definition and the differences between shapes. Pointedly, elemental features remain constant: amplitude $A$, wavelength $\lambda$, angle $\alpha$, horizontal length $l_h$ and inclined length $l_i$. Generally speaking, the height $A$ is reached through one, two or three jumps or steps by travelling the same distance in the horizontal axis. Particular geometrical values of the hierarchical profiles can be observed in Table \ref{tab_geometry}.

\begin{table}[h!]
	\centering
	\begin{tabular}{c|c|c|c|c|c}
		\toprule
		Configuration   & $d_1 [\text{mm}]$          & $d_2^x[\text{mm}]$           & $\alpha[\text{rad}]$ & $l_h[\text{mm}]$& $l_i[\text{mm}]$ \\ \midrule
		Trapezoidal     & $\lambda/8$                & $\lambda/4$                & $ \arctan\frac{4A}{\lambda}$  & $\lambda/2$ & $2\sqrt{(d_2^x)^2+A^2}$ \\ 
		\rule{0pt}{5ex}   Bi-Trapezoidal  & $\dfrac{\lambda-4d_2^x}{6}$  & $\dfrac{A/2}{\tan \alpha}$ & $ \arctan \frac{4A}{\lambda}$ & $\lambda/2$ & $4\sqrt{(d_2^x)^2+(A/2)^2}$\\ 
		\rule{0pt}{5ex}   Tri-Trapezoidal & $\dfrac{\lambda-6d_2^x}{8}$  & $\dfrac{A/3}{\tan \alpha}$ & $ \arctan\frac{4A}{\lambda}$  & $\lambda/2$ & $6\sqrt{(d_2^x)^2+(A/3)^2}$\\
		\bottomrule
	\end{tabular}
	\caption{Horizontal section length $d_1$, total horizontal and inclined section length, $l_h$ and $l_i$ respectively, and angle of the sloped sections $\alpha$ in trapezoidal, bi-trapezoidal and tri-trapezoidal configurations. $d_2^x$ stands for the horizontal component of the length $d_2$.}
	\label{tab_geometry}
\end{table}

For the FE analysis, 4-node plane-strain elements (type CPE4 in \texttt{ABAQUS}{\textsuperscript{\textregistered}} library) are used in the GFC, the nylon and the adhesive 2D modelling. Approximately 270k elements are employed to discretize the adherents and about 130 elements form every trapezium of the adhesive. Red arrows in Fig.~\ref{specimen_hier} represent the normal direction corresponding to each section of the motif.

The same boundary conditions than those used in previous Section are applied to the current DCB-like tests: a vertical displacement at the upper left end of the specimen while the lower left end is pinned. Likewise, the control algorithm of Section \ref{snapback} is employed in the simulations.

\begin{figure}[h!]
	\centering
	\includegraphics[width =0.7 \textwidth]{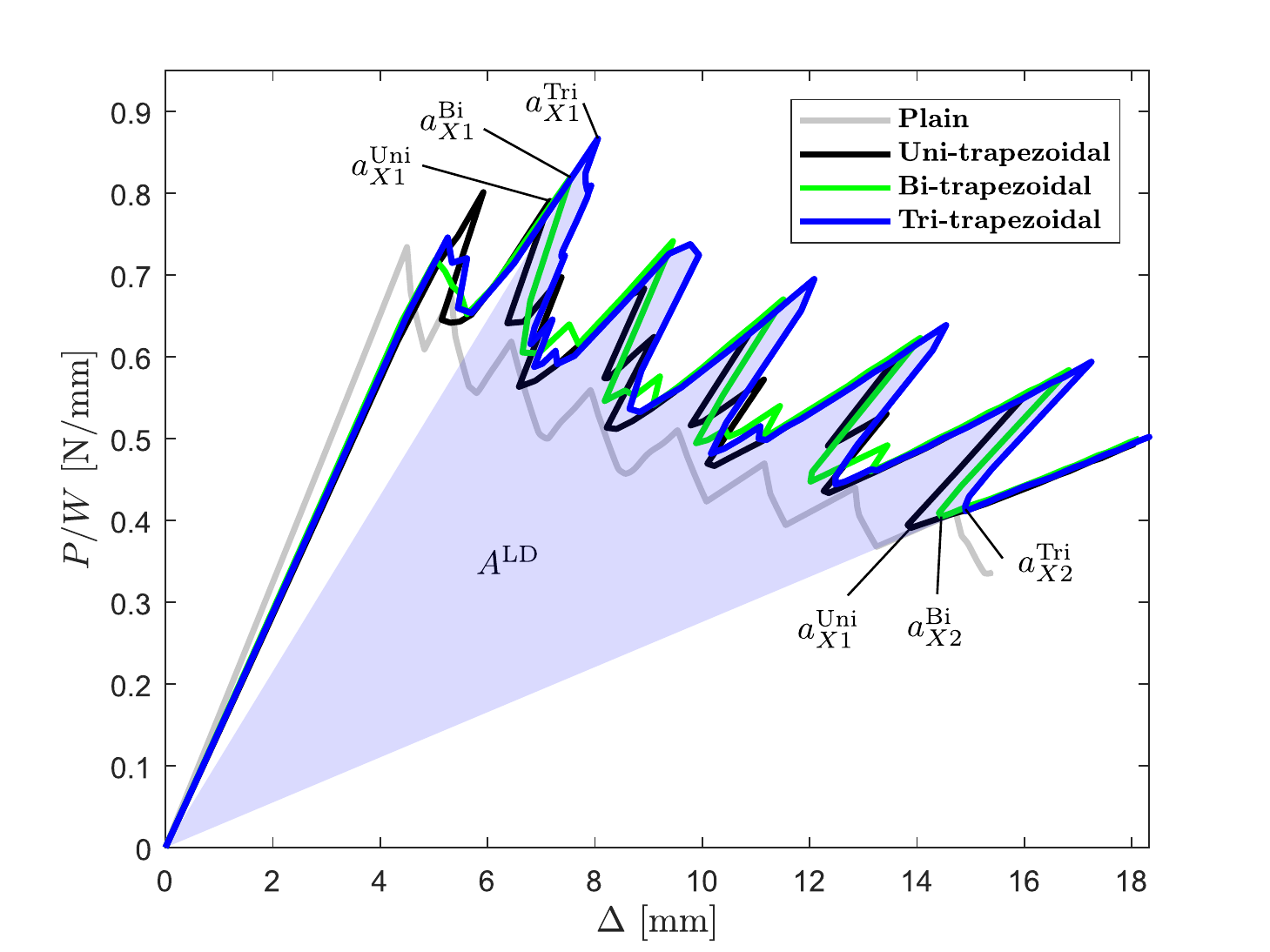}
	\caption{Load vs displacement curves of DCB tests with uni-trapezoidal, bi-trapezoidal and tri-trapezoidal interface profiles. $a_{X1}$ and $a_{X2}$ symbolise the initial and final lengths involved in the fracture characterisation and the shaded area $A^{\text{LD}}$ represents the area under the load-displacement curve used in Eq. \eqref{G_LD}.}
	\label{F_D}
\end{figure}

Considering previous aspects, numerical load-displacement curves of the uni-trapezoidal, bi-trapezoidal and tri-trapezoidal interface patterns, as well as the flat baseline scenario, can be observed in Fig.~\ref{F_D}. 

The behaviours of the three structured configurations are  similar to each other being characterized by: a first linear-elastic stage before damage onset and a region characterised by pronounced instabilities in the crack advance phase. Notwithstanding, a slight increase in the maximum load of the peaks can be appreciated with the hierarchical level. Additionally, the reference case presents unstable crack propagation despite of the flat interface, which leads to a saw-tooth force-displacement curve. It is worth mentioning that, if standard displacement control boundary conditions are applied, stabilization mechanisms of the solution would be required in order to achieved equilibrium solutions throughout the simulations. Nonetheless, the control algorithm discussed above allows the  computational convergence of the problem precluding  the use of any artificial damping energy.

\begin{figure}[h!]
	\centering
	\includegraphics[width =0.6 \textwidth]{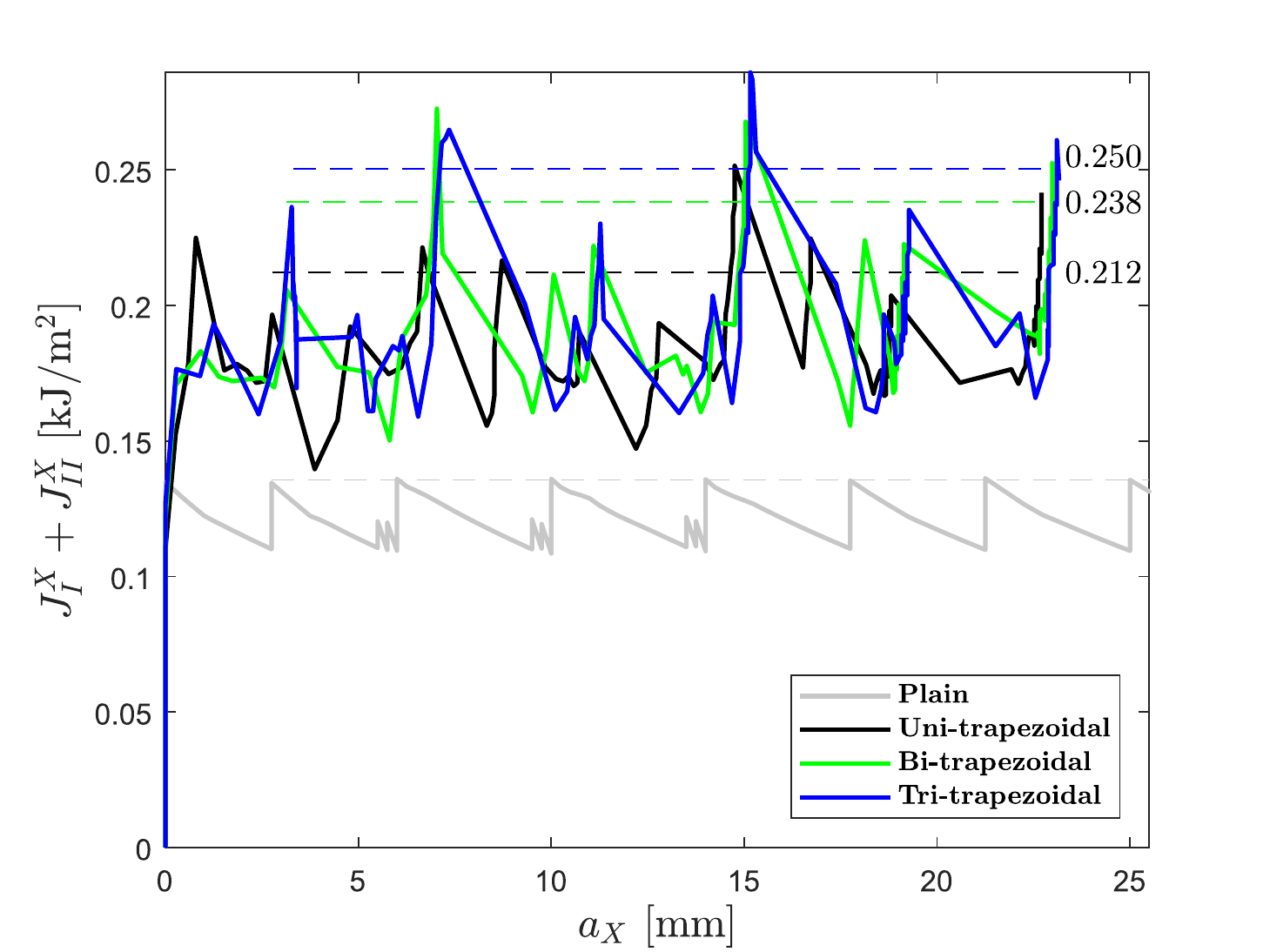}
	\caption{Effective J-Integral $J^X$ as a function of the effective crack length $a_X$ corresponding to the Double Cantilever Beam test with uni-, bi- and tri-trapezoidal crack paths.}
	\label{Jx_ax_hier}
\end{figure}

With reference to the qualitative response, the instabilities aforementioned can be also appreciated in Fig.~\ref{Jx_ax_hier}, where the effective J-Integral $J^X$, according to Eq. \eqref{J-Integral-Eq}, is represented as a function of the effective crack length $a_X$. In such graph, it is shown that the variability of the energy release rate is noticeable and the increase of the average fracture toughness (represented by the dashed lines and calculated by means of the peaks values) with respect to the level of arrangement. In this way, the tri-trapezoidal configuration achieves around 20\% of improvement with respect to the uni-trapezoidal interface and around 83\% with respect to the flat scenario.

\begin{figure}[h!]
	\centering
	\includegraphics[width =0.6 \textwidth]{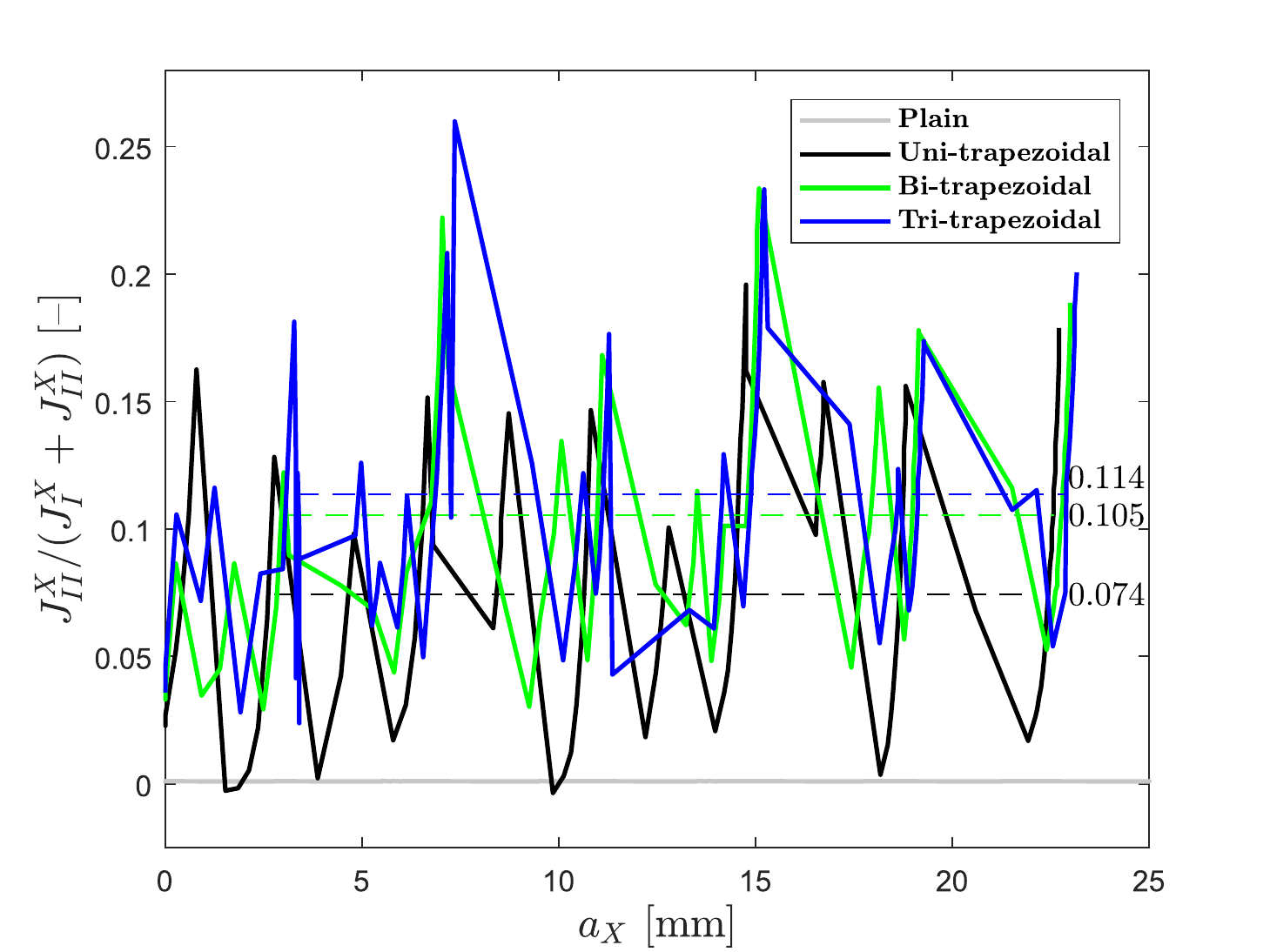}
	\caption{Effective mixed mode evolution $J_{II}^{X}/J_{T}^{X}$ at the crack tip as a function of the effective crack length $a_X$ corresponding to the Double Cantilever Beam test with uni-, bi- and tri-trapezoidal crack paths.}
	\label{GII_GT_hier}
\end{figure}

Regarding the fracture mixed mode of the three configurations, we can observe in Fig.~\ref{GII_GT_hier} the characteristic variability of the patterned interfaces, where the fluctuations become higher with the level of arrangement, that is, the tri-trapezoidal profile presents the highest variation of $J_{II}^X/J_{T}^X$. Furthermore, a minor increasing tendency of the mean value of the mixed mode during the test can be appreciated: $\frac{J_{II}^X}{J_{T}^X}\rvert_{\text{mean}}^{\text{Uni}} = 0.07$, $\frac{J_{II}^X}{J_{T}^X}\rvert_{\text{mean}}^{\text{Bi}} = 0.10$ and $\frac{J_{II}^X}{J_{T}^X}\rvert_{\text{mean}}^{\text{Tri}} = 0.11$.

Table \ref{G_J_mix} outlines the principal fracture energy values $G_{c}^{\text{LD}}$, $\bar{J}_c^{X}$ and the mean mixed mode $\frac{J_{II}^X}{J_{T}^X}\rvert_{\text{mean}}$ during the DCB test.

\begin{table}[h!]
	\centering
	\begin{tabular}{|c|c|c|c|}
		\hline
		\addlinespace[1ex]
		Configuration & $G_{c}^{\text{LD}}$ [J/m$^{2}$] & $\bar{J}_c^{X}$ [J/m$^{2}$] & $\frac{J_{II}^X}{J_{T}^X}\rvert_{\text{mean}}$ [-]\\ \hline
		Uni-Trapezoidal  & 195.9 & 212.3 & 0.074 \\ \hline
		Bi-Trapezoidal   & 205.1 & 238.1 & 0.105 \\ \hline
		Tri-Trapezoidal  & 210.5 & 250.5 & 0.114 \\ \hline
	\end{tabular}
	\caption{Energy release rate $G_{c}^{\text{LD}}$ obtained from load-displacement curves, mean effective J-Integral $\bar{J}_c^{X}$ and mean mixed mode $\frac{J_{II}^X}{J_{T}^X}\rvert_{\text{mean}}$ of the hierarchical trapezoidal interfaces in the DCB tests.}
	\label{G_J_mix}
\end{table}

The use of LEBIM to characterise the behaviour of the structured interfaces facilitates the comparison of the results determined from load-displacement and J-Integral methods. 
Such values differ between 8.4\% and 17.2\%. This difference arises from the energy calculation procedure: on the one hand, the area method implicitly implies an average of every point of the load-displacement curve (between the two crack lengths selected $a_{X1}$ and $a_{X2}$); on the other hand, the J-Integral average value is performed involving the peaks of the curve in Fig.~\ref{Jx_ax_hier}. As the mixed mode tendency aforementioned, the mean critical energy release rate ($G_{c}^{\text{LD}}$ or $\bar{J}_c^{X}$) increases with the level of arrangement, although such increment is small in comparison with the presence of the actual patterned interface. That is, with respect to the reference scenario (DCB test with flat interface, $G_{Ic}$), the simplest trapezoidal pattern implies an increment of the fracture toughness around the 56\%, whereas the trapezoidal profile with the highest level of arrangement (tri-trapezoidal), involves an increase of 84\%. Despite of the impact of the hierarchical arrangement is lower than the overall dimensions of the pattern (amplitude and wavelength) in the fracture properties, the growing tendency suggests that high levels of the arrangement of the geometry may be an interesting strategy to enhance the resistance of adhesively bonded joints.

\section{Conclusions}
\label{section4}

A comprehensive framework of computational interface modelling has been presented herein. The following three techniques have been summarised with the aim to overwhelm difficulties during the analysis of interfaces with complex geometry: (i) an innovative versatile model to calculate interface gaps under large displacement conditions, (ii) the Linear Elastic Brittle Interface Model able to describe the abrupt failure phenomena present in some joints, (iii) and a control algorithm to deal with instabilities result from the fracture mixed-mode variability along non-flat interface patterns. The first two methods have been embedded in a material user-subroutine \texttt{UMAT} of the software package \texttt{ABAQUS}{\textsuperscript{\textregistered}} whereas the latter one was performed by linking the applied boundary conditions with the crack tip opening employing auxiliary elements.

Aforementioned scheme was applied to delamination of composite laminates in a large range of mixed-mode fracture conditions: Double Cantilever Beam (DCB), Mixed Mode Bending (MMB) and End Notch Flexure (ENF) tests. Numerical results obtained from the Finite Element analysis were compared with experimental test available in the literature. The numerical-experimental correlation exhibits an excellent agreement and the employment of this interface modelling in structures involving a large variety of mixed-mode fracture conditions is justified.

The strategy proposed was exploited in a structured interface DCB Finite Element tests with different orders of hierarchical organization. In particular, uni-trapezoidal, bi-trapezoidal and tri-trapezoidal profiles were examined in the simulations. The load-displacement curves present analogous behaviours, developing a linear-elastic phase before damage appearance and consecutive saw-tooth responses during the crack propagation. It is worth mentioning that the higher order of the arrangement in the pattern the slightly larger fluctuations and higher maximum peak values are obtained. As load-displacement curves anticipate, energy release rate does not rely strongly on the hierarchical order considering the situations and the geometrical parameter suggested in this work. Then, a higher level of hierarchical arrangement may be needed to achieve a noticeable improvement in the interface fracture properties. Additionally, the shape and overall dimensions of the pattern may have more influence than the arrangement level.

The improvement of the present model can lead to interesting future research lines: 
\begin{itemize}
	\item LEBIM can be used with several damage criteria as described in \cite{Mantic}. Recently, Hutchinson and Suo and Quadratic criteria were used in a LEBIM implementation together with the Coupled Criterion of Finite Fracture Mechanics for the study of the fibre-matrix interface behaviour in \cite{MUNOZREJA2016267}. Obtained results were similar to each other for both criteria. The investigation of different damage criterion of interfaces (Hutchinson-Suo, power laws, etc) using LEBIM may widen the applicability of this tool in distinct scenarios, where differences on the mode mixity and the shape of the damaged area along the interface will appear.	
	\item The extension of this model to 3D applications in order to address interlaminar damage in intricate geometries.
	\item LEBIM could be compatible with fatigue behaviour or other environmental factors. Specifically, delamination of composite laminates were previously studied using a continuous distribution of linear elastic springs under cyclic loads \cite{BENNATI2006248,BENNATI201672}, so that LEBIM can be an appropriate tool to describe these type of events. Moreover, the standpoint developed herein for modelling the interface behaviour under finite deformation hypothesis can be combined with progressive damage theories so as to analyse fatigue loading under mixed mode conditions \cite{ZHANG2020102498}, as those presented in \cite{HOSSEINITOUDESHKY2020102480,ROCHA2020102493}. Such investigations	analyse the case of fatigue using elastic interfaces and/or Cohesive Zone Models. Moreover, the implementation of the LEBIM as user-defined element concern the possible integration  of additional features within the interface element such as fatigue \cite{DELBUSTO2017210}.
	\item The framework developed herein can be employed to model the behaviour of short fiber reinforced composites (SFRCs) at different scales of observation \cite{DEAN2019630,DEAN2016162,DEAN2016186,DEAN2017241}, being a matter that requires   comprehensive investigation activities.
\end{itemize}

\section*{Acknowledgements}
This study was supported by the Spanish Ministry of Science, Innovation and Universities and European Regional Development Fund (Project PGC2018-099197-B-I00), the Consejer\'ia de Econom\'ia y Conocimiento of the Junta de Andaluc\'ia (Spain) and Programa Operativo FEDER Andaluc\'a 2014-2020 for financial support under the contracts US-1265577, US-1266016 and AT17-5908-USE (Acciones de transferencia del conocimiento).


\bibliographystyle{model3-num-names}
\bibliography{garcia}

\newpage

\end{document}